\documentclass[prd,superscriptaddress,nofootinbib]{revtex4}
\usepackage{amsmath,amssymb,epsfig,color}
\usepackage[hypertex]{hyperref}
\usepackage{natbib,ifthen}
\usepackage{bm}

\begin{document}

\newcommand{\fixme}[1]{{\textbf{Fixme: #1}}}
\newcommand{\AC}[1]{{\textcolor{red}{#1}}}
\newcommand{\ACtwo}[1]{\textcolor{green}{\textbf{AC}: #1}}
\newcommand{\detD}{{\det\!\cld}}
\newcommand{\clh}{\mathcal{H}}
\newcommand{\ud}{{\rm d}}
\renewcommand{\eprint}[1]{\href{http://arxiv.org/abs/#1}{#1}}
\newcommand{\adsurl}[1]{\href{#1}{ADS}}
\newcommand{\ISBN}[1]{\href{http://cosmologist.info/ISBN/#1}{ISBN: #1}}
\newcommand{\jcap}{J.\ Cosmol.\ Astropart.\ Phys.}
\newcommand{\mnras}{Mon.\ Not.\ R.\ Astron.\ Soc.}
\newcommand{\progress}{Rep.\ Prog.\ Phys.}
\newcommand{\prlett}{Phys.\ Rev.\ Lett.}
\newcommand{\aap}{A\&A}
\newcommand{\aapr}{A\&A Rev.}
\newcommand{\vort}{\varpi}
\newcommand\ba{\begin{eqnarray}}
\newcommand\ea{\end{eqnarray}}
\newcommand\be{\begin{equation}}
\newcommand\ee{\end{equation}}
\newcommand\lagrange{{\cal L}}
\newcommand\cll{{\cal L}}
\newcommand\cln{{\cal N}}
\newcommand\clx{{\cal X}}
\newcommand\clz{{\cal Z}}
\newcommand\clv{{\cal V}}
\newcommand\cld{{\cal D}}
\newcommand\clt{{\cal T}}

\newcommand{\ThreeJSymbol}[6]{\begin{pmatrix}
#1 & #3 & #5 \\
#2 & #4 & #6
 \end{pmatrix}}

\newcommand\clo{{\cal O}}
\newcommand{\cla}{{\cal A}}
\newcommand{\clp}{{\cal P}}
\newcommand{\clr}{{\cal R}}
\newcommand{\uD}{{\mathrm{D}}}
\newcommand{\calE}{{\cal E}}
\newcommand{\calB}{{\cal B}}
\newcommand{\curl}{\,\mbox{curl}\,}
\newcommand\del{\nabla}
\newcommand\Tr{{\rm Tr}}
\newcommand\half{{\frac{1}{2}}}
\newcommand\fourth{{1\over 8}}
\newcommand\bibi{\bibitem}
\newcommand{\kf}{\beta}
\newcommand{\kfprod}{\alpha}
\newcommand\calS{{\cal S}}
\renewcommand\H{{\cal H}}
\newcommand\K{{\rm K}}
\newcommand\mK{{\rm mK}}
\newcommand\km{{\rm km}}
\newcommand\synch{\text{syn}}
\newcommand\opacity{\tau_c^{-1}}

\newcommand{\Psil}{\Psi_l}
\newcommand{\bsigma}{{\bar{\sigma}}}
\newcommand{\bI}{\bar{I}}
\newcommand{\bq}{\bar{q}}
\newcommand{\bv}{\bar{v}}
\renewcommand\P{{\cal P}}
\newcommand{\numfrac}[2]{{\textstyle \frac{#1}{#2}}}

\newcommand{\la}{\langle}
\newcommand{\ra}{\rangle}

\newcommand{\Omtot}{\Omega_{\mathrm{tot}}}
\newcommand\xx{\mbox{\boldmath $x$}}
\newcommand{\phpr} {\phi`}
\newcommand{\gam}{\gamma_{ij}}
\newcommand{\sqgam}{\sqrt{\gamma}}
\newcommand{\delk}{\Delta+3{\K}}
\newcommand{\dph}{\delta\phi}
\newcommand{\om} {\Omega}
\newcommand{\dom}{\delta^{(3)}\left(\Omega\right)}
\newcommand{\rar}{\rightarrow}
\newcommand{\Rar}{\Rightarrow}
\newcommand\gsim{ \lower .75ex \hbox{$\sim$} \llap{\raise .27ex \hbox{$>$}} }
\newcommand\lsim{ \lower .75ex \hbox{$\sim$} \llap{\raise .27ex \hbox{$<$}} }
\newcommand\bigdot[1] {\stackrel{\mbox{{\huge .}}}{#1}}
\newcommand\bigddot[1] {\stackrel{\mbox{{\huge ..}}}{#1}}
\newcommand{\Mpc}{\text{Mpc}}
\newcommand{\Al}{{A_l}}
\newcommand{\Bl}{{B_l}}
\newcommand{\eAl}{e^\Al}
\newcommand{\ix}{{(i)}}
\newcommand{\ixp}{{(i+1)}}
\renewcommand{\k}{\beta}
% Derivatives
\newcommand{\HD}{\mathrm{D}}

\newcommand{\nonflat}[1]{#1}
\newcommand{\Cgl}{C_{\text{gl}}}
\newcommand{\Cgltwo}{C_{\text{gl},2}}
\newcommand{\He}{{\rm{He}}}
\newcommand{\Mhz}{{\rm MHz}}
\newcommand{\vx}{{\mathbf{x}}}
\newcommand{\ve}{{\mathbf{e}}}
\newcommand{\vv}{{\mathbf{v}}}
\newcommand{\vk}{{\mathbf{k}}}
\newcommand{\vl}{{\mathbf{l}}}
\newcommand{\vn}{{\mathbf{n}}}

\newcommand{\vnhat}{{\hat{\mathbf{n}}}}
\newcommand{\vkhat}{{\hat{\mathbf{k}}}}
\newcommand{\taueps}{{\tau_\epsilon}}

\newcommand{\vgrad}{{\mathbf{\nabla}}}
\newcommand{\fbarln}{\bar{f}_{,\ln\epsilon}(\epsilon)}

%%%%%%%%%%%%%%%%%%%%%%%%%%%%%%%%%%%%%%%%%%%%%%%%

\title{Detecting the polarization induced by scattering of the microwave background quadrupole in galaxy clusters}

\author{Alex Hall}
\email{ahall@roe.ac.uk}
\affiliation{Institute for Astronomy, University of Edinburgh, Royal Observatory, Blackford Hill, Edinburgh, EH9 3HJ, U.K.}
\affiliation{Institute of Astronomy and Kavli Institute for Cosmology, Madingley Road, Cambridge, CB3 0HA, U.K.}

\author{Anthony Challinor}
\affiliation{Institute of Astronomy and Kavli Institute for Cosmology, Madingley Road, Cambridge, CB3 0HA, U.K.}
 \affiliation{DAMTP, Centre
for Mathematical Sciences, Wilberforce Road, Cambridge CB3 0WA, U.K.}

%\date{\today}
%\pubyear{2005}

%\label{firstpage}

\begin{abstract}
We analyse the feasibility of detecting the polarization of the CMB caused by scattering of the remote temperature quadrupole by galaxy clusters with forthcoming CMB polarization surveys. For low-redshift clusters, the signal is strongly correlated with the local large-scale temperature and polarization anisotropies, and the best prospect for detecting the cluster signal is via cross-correlation. For high-redshift clusters, the correlation with the local temperature is weaker and the power in the uncorrelated component of the cluster polarization can be used to enhance detection. We derive linear and quadratic maximum-likelihood estimators for these cases, and forecast signal-to-noise values for the SZ surveys of a \emph{Planck}-like mission and SPTPol. Our estimators represent an optimal `stacking' analysis of the polarization from clusters. We find that the detectability of the effect is sensitive to the cluster gas density distribution, as well as the telescope resolution, cluster redshift distribution, and sky coverage. We find that the effect is too small to be detected in current and near-future SZ surveys without dedicated polarization follow-up, and that an r.m.s.\ noise on the Stokes parameters of roughly 1~$\mu$K-arcmin for each cluster field is required for a 2\,$\sigma$ detection, assuming roughly 550 clusters are observed. We show that ACTPol is in a better position to observe the effect than SPTPol due to its advantageous survey location on the sky, and we discuss the novel spatial dependence of the signal. We discuss and quantify potential biases from the kinetic part of the signal caused by the relative motion of the cluster with respect to the CMB, and from the background CMB polarization behind the cluster, discussing ways in which these biases might be mitigated. Our formalism should be important for next-generation CMB polarization missions, which we argue will be able to measure this effect with high signal-to-noise. This will allow for an important consistency test of the $\Lambda \mathrm{CDM}$ model on scales that are inaccessible to other probes.

%However, we argue that a moderate increase in the integration time per cluster will allow for a detection
\end{abstract}

\maketitle

\section{Introduction}
\label{sec:intro}

Inverse-Compton scattering of cosmic microwave background (CMB) photons by free electrons in galaxy clusters gives rise to spectral distortions, the Sunyaev-Zeldovich (SZ) effect \citep{1972CoASP...4..173S}. This effect has been used extensively to identify clusters from maps of the CMB temperature anisotropies, with current surveys able to detect hundreds of clusters \citep{2013arXiv1301.0816H,2013ApJ...763..127R,2013arXiv1303.5089P}.

Scattering also imparts linear polarization into the CMB, with an amplitude proportional to the radiation quadrupole in the rest-frame of the scattering electron \citep{1980SvAL....6..285Z, 1980MNRAS.190..413S, 1981ASPRv...1....1S}. The effect in clusters has not yet been observed, but its detection could potentially allow one to measure the CMB quadrupole at the spacetime location of the cluster, providing an indirect snapshot of the Universe within our past light-cone \citep{1997PhRvD..56.4511K}. We know that our local quadrupole is low compared to the best-fit $\Lambda$CDM model, so this technique could be used to compare our value with the value in other parts of the Universe. It can also be used to make measurements of the large-scale density field that are independent of those provided by other cosmological probes \citep{2006PhRvD..73l3517B}.

With redshift information from optical observations, the cluster polarization could be used to measure the time-evolution of the quadrupole, which is sensitive to late-time structure formation through the integrated Sachs-Wolfe (ISW) effect, hence providing a window on dark energy \citep{2003PhRvD..67f3505C,2005bmri.conf..309C}.

In this work, we discuss how this signal can be first \emph{detected} in the optimal way. Current and near-future CMB polarization experiments such as \emph{Planck}, ACTPol and SPTPol, as well as the proposed next-generation mission PRISM \citep{2013arXiv1306.2259P}, will provide high-quality measurements of polarization in the microwave sky, so a detailed study appears timely. We construct optimal estimators for the effect, quantify the impact of the major systematic errors and propose techniques to mitigate them, and forecast signal-to-noise estimates for current and future surveys. Our analysis differs from previous studies in that we focus on the detection of the signal rather than its statistics, which are well understood on the cosmological scales relevant for CMB experiments \citep{2004PhRvD..70f3504P,2005NewA...10..417A, 2006PhRvD..73l3517B,2012ApJ...757...44R}. Detection of the cluster polarization signal with the expected amplitude would be a non-trivial test of the cosmological model on the largest observable scales.

However, there are several complications that must be overcome before the primordial quadrupole can be measured. Firstly, if the cluster is in relative motion with respect to the CMB rest-frame\footnote{Hereafter, the term `CMB rest-frame' will be used to refer to the frame in which the CMB dipole vanishes.}, Doppler and aberration effects will produce polarization even in the absence of a primordial quadrupole \citep{1980MNRAS.190..413S, 1999MNRAS.310..765S, 2000MNRAS.312..159C}. The dominant effect for typical cluster velocities and gas temperatures is the transformation of the CMB rest-frame monopole into a quadrupole in the electron rest-frame. For unpolarized incident radiation, this gives linear polarization in the plane defined by the quadrupole, and hence the observed polarization is proportional (at leading order) to the square of the \emph{transverse} velocity of the cluster with respect to the line of sight. We shall refer to this effect as the kinetic part of the signal. 

The kinetic part has a unique frequency-dependence, a consequence of the Lorentz transformation between the electron rest-frame and the CMB rest-frame. In contrast, the primordial part has the same frequency-dependence as the temperature quadrupole, i.e., the first derivative of a blackbody. This is a consequence of Thomson scattering not altering the energy of the incident photon (for higher-order and relativistic corrections, see \citep{2000MNRAS.312..159C,2000ApJ...533..588I}). The opposite is true for the usual SZ effect, where scattering of the primordial CMB by the distribution of electrons in the intra-cluster medium (ICM; the thermal SZ effect) \emph{does} give rise to a spectral distortion, which is how the cluster may be identified. However, the part of the signal due to the cluster's bulk velocity (the kinetic SZ effect) has the \emph{same} frequency-dependence as the incident radiation. Moreover, the amplitude of this effect is proportional to the \emph{radial} cluster bulk velocity. Detections of this effect have recently been made statistically by the Atacama Cosmology Telescope \citep{2012PhRvL.109d1101H}, and directly towards a massive cluster with Bolocam measurements~\citep{2013ApJ...778...52S}.

Fortunately, for typical cluster velocities ($v \sim \mathcal{O}(10^{2})$ $ \km \, \rm s^{-1}$) the amplitude of the kinetic polarization signal is roughly a factor of 10 smaller than the primordial part at typical observing frequencies. In addition, the unique frequency dependence of this effect should allow it to be distinguished from the primordial signal if multi-frequency data are available.

An additional systematic comes from the background CMB polarization, which has the same frequency-dependence as the cluster quadrupole signal, but a different correlation structure. This could potentially bias the measurement if not properly accounted for. This background is also correlated with the temperature quadrupole that generates the signal itself, which further complicates the analysis. Compton scattering also has a dependence on the incident polarization leading to an additional correlation of the signal with the background CMB.

Since clusters are optically thin to Compton scattering, the generated polarization is linearly proportional to the optical depth through the cluster. To estimate the amplitude of the primordial signal it is necessary to know the cluster optical depth profile. This may be extracted from X-ray surface brightness observations if the temperature profile is known, which requires high-resolution spectroscopic X-ray measurements (see e.g. \citet{2006ApJ...640..691V}). However, X-ray measurements are only available beyond the virial radius for a few systems, leading to uncertainty in the gas profile and hence optical depths in the cluster outskirts. At larger radii we may appeal to simulations, but these show a high degree of scatter in the gas profile of the outer region, dependent on cluster mass \citep{2007MNRAS.380..877S}. Thus our inferences of optical depths beyond a few virial radii are likely to be unreliable. In principle, relativistic corrections to the thermal SZ effect could help to constrain optical depths without the use of X-ray information, but these corrections are likely to be small for most of the clusters we consider \citep{1998ApJ...499....1C,1998ApJ...502....7I}.

Further potential systematics are provided by polarized foreground contamination, as in conventional CMB polarization analyses, and polarized emission intrinsic to the cluster itself. We will not attempt to address foregrounds in this work, since the techniques required to remove them are the same as in conventional analyses of small-scale CMB anisotropies (e.g.,~\cite{2009AIPC.1141..222D}). We briefly discuss the impact of intrinsic polarization in Sec.~\ref{sec:conc}.

Despite these obstacles, the signal we are trying to detect has a well-understood scale-dependence. At low redshift, the signal is proportional to the local quadrupole, and hence is correlated over very large angular scales. The signal from high-redshift clusters contains more angular structure when observed today due to free-streaming, but for typical survey depths the large-scale nature of the signal is preserved. An SZ polarization survey then amounts to extracting noisy, sparse samples from a 3D field with a large correlation length. In addition, the signal is correlated with the local temperature and polarization anisotropies. These properties must be used in an optimal way to overcome the problems alluded to above.

The plan of this paper is as follows. In Sec.~\ref{sec:physics} we discuss the physics of the signal we seek to measure, and discuss its statistics. In Sec.~\ref{sec:like} we write down the likelihood function for the signal and derive the maximum-likelihood estimator for the amplitude of the signal for low-redshift cluster surveys.
In Sec.~\ref{sec:sn} we discuss our survey and cluster catalogue assumptions, and forecast $S/N$ estimates. We discuss potential biases in Secs.~\ref{sec:K-bias} and~\ref{sec:E-bias}. In Sec.~\ref{sec:hi-z} we generalize our estimator to high-redshift cluster surveys, and we conclude in Sec.~\ref{sec:conc}. We derive the harmonic-space representations of the cluster quadrupole signal in Appendix~\ref{app:harm} and the kinetic signal in Appendix~\ref{app:vlm}.

The fiducial model we assume in calculating the statistics of the signal is a flat $\Lambda$CDM universe with parameters \{$\Omega_b h^2, \Omega_c h^2 , h, \tau, 10^9 A_s, n_s$\} given by \{$0.0223,0.112,0.7,0.09, 2.1, 0.96$\}. We neglect the effects of massive neutrinos and dynamical dark energy, and consider only scalar adiabatic initial conditions evolved with linear perturbation theory.

\section{Polarization observed towards galaxy clusters}
\label{sec:physics}

\subsection{Thomson scattering of the CMB}

%In this section we summarize the calculation of the polarization signal we seek to measure. We will consider the single scattering of a CMB photon off a free electron in the electron's rest-frame. We will work in the Thomson limit, neglecting the effects of electron recoil. In this limit, the energy of the photon is not altered by scattering.
%
In this section, we review the calculation of the generation of polarization from Thomson scattering (see, e.g.,~\citep{1997PhRvD..56..596H}). We will consider the single scattering of a CMB photon off a free electron in the electron's rest-frame, neglecting the effects of electron recoil. 

Polarization is most easily handled by use of the Stokes parameters $Q$ and $U$. These are elements of a symmetric trace-free tensor $\mathcal{P}_{ab}$, which is defined in the two-space orthogonal to both the propagation direction and the observer's four-velocity. Specifically we have 
\begin{equation}
\mathcal{P}_{11} = Q/2, \; \; \mathcal{P}_{12} = \mathcal{P}_{21} = U/2, \; \; \mathcal{P}_{22} = -Q/2,
\end{equation}
where the indices refer to components on an orthonormal tetrad with 3-direction aligned with the propagation direction.

We will simplify the scattering geometry by setting up a spherical coordinate system centred on the scattering electron. We can define a local orthonormal right-handed basis on the sphere with the standard basis vectors $\{\hat{\mathbf{x}},\hat{\mathbf{y}},\hat{\mathbf{z}}\}=\{\hat{\bm{\theta}},\hat{\bm{\phi}},\ve\}$, where $\ve$ is the radial unit vector giving the direction of photon propagation.

A quantity ${}_{s}\eta$ has spin $s$ if, under the transformation
\begin{equation}
\hat{\bm{\theta}} + i\hat{\bm{\phi}} \rightarrow (\hat{\bm{\theta}} + i\hat{\bm{\phi}})e^{i\gamma},
\end{equation}
we have ${}_{s}\eta \rightarrow e^{i s\gamma}{}_{s}\eta$. In our coordinate system, the complex polarization $Q+iU$ thus has spin $2$, and $Q-iU$ has spin $-2$.

The photon is incident along a direction $\ve_1$, gets scattered through an angle $\beta$ and leaves along a direction $\ve_2$, with $\ve_1 \cdot \ve_2 = \cos\beta$. We denote the increment of physical distance along the line of sight by $\mathrm{d}l$.
We set up polarization bases at $\ve_1$ and $\ve_2$ that have their local $x$-axes in the scattering plane and local $y$-axes perpendicular to it. We denote with an overbar the polarization defined with respect to these bases. For unpolarized incident radiation temperature $T(\vnhat_1)$ the outgoing photon has $\mathrm{d} \bar{U}=0$ by symmetry, and 
\begin{equation}
\mathrm{d} \bar{Q}(\ve_2) =- \frac{3}{16\pi}n_e \sigma_{\mathrm{T}} \mathrm{d} l  \sin^2 \beta T(\ve_1)\, \mathrm{d} \ve_1,
\end{equation}
where $n_e$ is the free electron density of the medium, and $\sigma_{\mathrm{T}}$ is the Thomson cross-section (e.g.,~\citep{1997PhRvD..56..596H}).

Define the angle $\gamma_2$ as the angle required to rotate (in a right-handed sense)
the $\{\hat{\bm{\theta}},\hat{\bm{\phi}}\}$ basis at $\ve_2$ onto the scattering-plane basis there. Let the angle $\gamma_1$ be the corresponding quantity at $\ve_1$. The angles $\{\gamma_1,\beta,-\gamma_2\}$ form a set of Euler angles that rotate the polar basis at $\ve_1$ onto that at $\ve_2$.

Performing the reverse rotation at $\ve_2$ from the scattering plane onto the spherical coordinate basis, we have 
\begin{equation}
\mathrm{d}(Q \pm  iU) = -\frac{3}{16\pi}n_e \sigma_{\mathrm{T}} \mathrm{d} l \, \sin^2 \beta \,e^{\pm 2i\gamma_2} \,T(\ve_1) \,\mathrm{d} \ve_1 .
\end{equation}
We can express $\sin^2 \beta$ in terms of the elements of a Wigner $D$-matrix \citep{QTAM}
\begin{equation}
\sin^2 \beta = \sqrt{\frac{8}{3}}D^2_{0 \pm 2}(\gamma_1,\beta,-\gamma_2)e^{\mp 2i\gamma_2}.
\end{equation}
We now make use of the addition theorem for spin-weighted spherical harmonics,
\begin{equation}
D^l_{s s'}(\gamma_1, \beta, -\gamma_2) = \sum_m \frac{4 \pi}{2l + 1} {}_s Y^*_{lm}(\ve_1) {}_{s'} Y_{lm}(\ve_2).
\end{equation}
Using this, and the relation between Wigner $D$ matrix elements and spin harmonics
\begin{equation}
D^l_{-m s}(\phi, \theta, 0) = (-1)^m \sqrt{\frac{4 \pi}{2l+1}} {}_s Y_{lm}(\ve),
\end{equation}
we have for the outgoing polarization along $\ve_2$
\begin{equation}
\mathrm{d}(Q\pm iU) = -\frac{3}{16\pi}n_e \sigma_{\mathrm{T}} \mathrm{d} l \,T(\ve_1)\,\mathrm{d} \ve_1 \sqrt{\frac{8}{3}} \sum_m \frac{4\pi}{5} Y^*_{2m}(\ve_1)\,{}_{\pm 2}Y_{2m}(\ve_2).
\end{equation}
This has the form of a spin 2 expansion, confirming that $(Q+iU)$ is spin 2 with our current conventions. Integrating over all incoming directions $\ve_1$ we have
\begin{equation}
\label{eq:pol_out}
\mathrm{d}(Q \pm iU) = -\frac{1}{10} \mathrm{d} \tau \sum_m \sqrt{6} a_{2m} \,{}_{\pm 2}Y_{2m}(\ve_2),
\end{equation}
where we have expanded the temperature of the incident distribution in spherical harmonics $T(\ve) = \sum_{lm} a_{lm}Y_{lm}(\ve)$, and identified $n_e \sigma_{\mathrm{T}} \mathrm{d} l$ with an increment in the optical depth $\mathrm{d} \tau$. Equation~\eqref{eq:pol_out} should be compared with the general expansion of the polarization in spin-$\pm 2$ harmonics
\begin{equation}
(Q \pm iU)(\ve) = \sum_{lm}(E_{lm} \pm iB_{lm}){}_{\pm 2} Y_{lm}(\ve).
\end{equation}
Under a parity transformation, $(Q \pm iU)(\ve) \rightarrow (Q \mp iU)(-\ve)$, so $E_{lm}$ has $(-1)^l$ (electric) parity and $B_{lm}$ has $(-1)^{l+1}$ (magnetic) parity. The temperature transforms as $T(\ve) \rightarrow T(-\ve)$, so $a_{lm}$ has electric parity. Therefore the polarization is locally of the purely electric quadrupole type. Equation~\eqref{eq:pol_out} also makes it clear that polarization from Thomson scattering is generated by quadrupolar anisotropies in the incident radiation field \citep{Chandra}.

We now need to relate the polarization carried away by the photon in direction $\ve_2$ to the polarization measured by the observer. We will set up a new spherical coordinate system centred on the observer, who observes the radiation along a line of sight $\vnhat$. A local right-handed basis can be set up on the sphere with the 3-direction aligned with the propagation direction $\ve = -\vnhat$ if we choose the local $\{x,y\}$ basis as $\{\hat{\bm{\theta}},-\hat{\bm{\phi}}\}$. However, some care must be taken as $Q+iU$ on this new basis is spin $-2$ \citep{2004MNRAS.350..914C}. The observed polarization is obtained by performing a parity transformation on Eq.~\eqref{eq:pol_out}, since the new basis has had its local 3-direction reversed.

Since the cluster is optically thin, we can neglect any further scattering of the photon. Integrating through the cluster gives the observed complex polarization 
\begin{equation}
\label{eq:pol_in}
(Q \pm iU) (\vnhat) = -\frac{\sqrt{6}}{10}\tau(\vnhat)\sum_{m=-2}^{2}a_{2m}(\mathbf{r}){}_{\mp 2}Y_{2m}(\vnhat),
\end{equation}
where $a_{2m}(\mathbf{r})$ denotes the CMB quadrupole at the location of the cluster. This expression matches Eq.~(7) of \citep{2006PhRvD..73l3517B}, up to an arbitrary overall sign. Note that for polarized incident radiation, $a_{2m}$ in the above expression should be replaced by $a_{2m} - \sqrt{6}E_{2m}$ where $E_{2m}$ is the $E$-mode quadrupole. However, the correction is very small for scattering well after recombination.

Equation~\eqref{eq:pol_in} tells us that the observed polarization is purely quadrupolar for a very low-redshift cluster. However, to calculate the angular structure of the signal due to scattering at higher redshift, we must expand $a_{2m}(\mathbf{r}){}_{\mp 2}Y_{2m}(\vnhat)$ in spin $\mp 2$ spherical harmonics. The calculation appears in \citep{1997PhRvD..56..596H,2006PhRvD..73l3517B}, so we will not reproduce it here. The result is 
\begin{equation}
\label{eq:harm_exp}
(Q \pm iU) (\vnhat)\tau^{-1}(\vnhat) = \sum_{lm} p_{lm}(r){}_{\mp 2}Y_{lm}(\vnhat),
\end{equation}
where $r = |\mathbf{r}|$ and the coefficients are given by
\begin{equation}
\label{eq:plm}
p_{lm}(r) = -i^l 3\pi \sqrt{\frac{(l+2)!}{(l-2)!}} \int \frac{\mathrm{d}^3 \mathbf{k}}{(2\pi)^{3/2}} \, \frac{j_l(kr)}{(kr)^2} \Delta_2(k;r) 
\Phi(\mathbf{k}) Y_{lm}^*(\hat{\mathbf{k}}) .
% p_{lm}(r) = -i^l\sqrt{\frac{3}{40\pi}}\int \frac{\mathrm{d}^3 \mathbf{k}}{(2\pi)^{3/2}} \, \Delta_2(k;r) \Phi(\mathbf{k}) F_l(kr) Y_{lm}^*(\hat{\mathbf{k}}),
\end{equation}
Here, $\Delta_2(k;r)$ is the transfer function relating the gravitational potential in matter-domination, $\Phi(\mathbf{k})$, to the temperature quadrupole $a_{2m}(\mathbf{k})$ at conformal time $\eta=\eta_0-r$, given by
\begin{equation}
\Delta_2(k;r) = \frac{1}{3}j_2[k(\eta - \eta_*)] + 2\int_{\eta_*}^{\eta}\mathrm{d}\eta' \, j_2[k(\eta - \eta')]\frac{\partial}{\partial \eta'}\left[\frac{D(\eta')}{a(\eta')}\right].
\end{equation}
The conformal time today (where $a=1$) is $\eta_0$ and $\eta_*$ is the conformal time at last scattering. The growth factor of the comoving-gauge matter perturbation, $D(a)$, is normalized to unity at high redshift in matter domination. We have assumed that $\Delta_2(k;r)$ contains contributions from the Sachs-Wolfe effect and the ISW effect only, i.e., we neglect the Doppler term whose contribution to the quadrupole is sub-dominant. It is easy to check from Eq.~\eqref{eq:plm} that $p_{lm}^* = (-1)^m p_{l-m}$, and  that the projection of the quadrupole preserves the $E$-mode nature of the signal.
% The geometric kernel in Eq.~\eqref{eq:plm} is 
% %
% \begin{equation}
% F_l(kr) = \sqrt{20\pi}\sum_{\lambda = l-2,l,l+2} (-1)^{(\lambda-l)/2}(2\lambda +1) \ThreeJSymbol{2}{2}{l}{-2}{\lambda}{0} \ThreeJSymbol{2}{0}{l}{0}{\lambda}{0} j_{\lambda}(kr) = - \sqrt{20\pi} \sqrt{\frac{3}{8}\frac{(l+2)!}{(l-2)!}}\frac{j_l(kr)}{(kr)^2}.
% \end{equation}
% %

\subsection{Correlation functions}

We assume that the temperature quadrupole can be calculated within linear theory, which should be a very accurate approximation. We also assume that the initial fluctuations are Gaussian and that temperature anisotropies are generated by scalar curvature perturbations.

In harmonic space, we can define the angular power spectra of the polarization field $p_{lm}(r)$, as well as its cross-correlation with the local CMB temperature anisotropies $a_{lm}$ and background CMB polarization anisotropies $E_{lm}$:
\begin{equation}
\label{eq:xi}
\langle p_{lm}(r)p^*_{l'm'}(r') \rangle = \delta_{ll'}\delta_{mm'}\xi_l(r,r'),
\end{equation}
\begin{equation}
\label{eq:zetaT}
\langle p_{lm}(r)a^*_{l'm'} \rangle = \delta_{ll'}\delta_{mm'}\zeta^T_l(r),
\end{equation}
\begin{equation}
\label{eq:zetaE}
\langle p_{lm}(r)E^*_{l'm'} \rangle = \delta_{ll'}\delta_{mm'}\zeta^E_l(r),
\end{equation}
\begin{equation}
\langle a_{lm}a^*_{l'm'} \rangle = \delta_{ll'}\delta_{mm'}C^{TT}_l,
\end{equation}
\begin{equation}
\langle E_{lm}E^*_{l'm'} \rangle = \delta_{ll'}\delta_{mm'}C^{EE}_l,
\end{equation}
\begin{equation}
\langle a_{lm}E^*_{l'm'} \rangle = \delta_{ll'}\delta_{mm'}C^{TE}_l.
\end{equation}
Note that parity invariance in the mean ensures vanishing correlation of the cluster polarization signal with $B$-mode polarization.

In real space, polarization correlation functions can be defined in a way that makes them independent of the coordinate system used to define the Stokes parameters. Recall that an overbar denotes a quantity evaluated on the `physical' basis defined by the geodesic connecting the two points of interest on the sphere. We may then define the correlation functions \citep{1997PhRvD..55.7368K, 1999IJMPD...8...61N}
\begin{equation}
\label{eq:corr_m}
\Xi^{-}(\beta,r,r') \equiv \langle \bar{p}(\vnhat_1; r)\bar{p}(\vnhat_2; r') \rangle = \sum_l \frac{2l+1}{4\pi}\xi_l(r,r')d^l_{2-2}(\beta),
\end{equation}
\begin{equation}
\label{eq:corr_p}
\Xi^{+}(\beta,r,r') \equiv \langle \bar{p}^*(\vnhat_1; r)\bar{p}(\vnhat_2; r') \rangle = \sum_l \frac{2l+1}{4\pi}\xi_l(r,r')d^l_{22}(\beta),
\end{equation}
\begin{equation}
\langle \bar{p}(\vnhat_1; r)T(\vnhat_2) \rangle = \sum_l \frac{2l+1}{4\pi}\zeta^T_l(r)d^l_{20}(\beta),
\end{equation}
where $d^l_{mm'}$ are elements of Wigner reduced matrices, and we have defined $p=(Q+iU)/\tau$. Equations~\eqref{eq:corr_m} and \eqref{eq:corr_p} correct Eqs.~(21) and (22) of \citep{2006PhRvD..73l3517B}, which did not properly account for spin-2 nature of polarization. The measured polarizations are given by the rotated quantities
\begin{equation}
\label{eq:p_rot1}
p(\vnhat_1)=e^{-2i\gamma_1}\bar{p}(\vnhat_1),
\end{equation}
\begin{equation}
\label{eq:p_rot2}
p(\vnhat_2)=e^{-2i\gamma_2}\bar{p}(\vnhat_2).
\end{equation}
The final ingredient required to calculate these two-point functions is the primordial power spectrum of $\Phi$. This is given by
\begin{equation}
\langle \Phi(\mathbf{k}) \Phi^*(\mathbf{k}') \rangle = P_{\Phi}(k)\delta^{(3)}(\mathbf{k}-\mathbf{k}'),
\end{equation}
with $P_{\Phi}(k) \propto k^{n_s-4}$ where the scalar spectral index $n_s \approx 1$.

\begin{figure}
\centering
\includegraphics[width=0.6\textwidth]{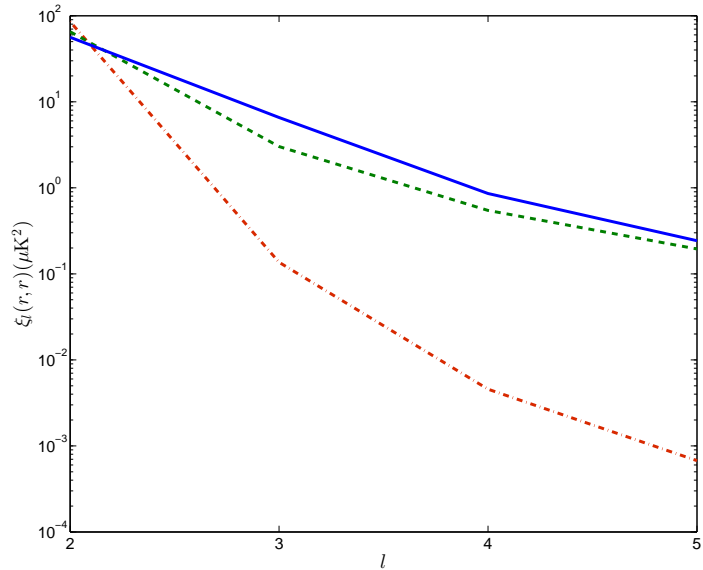}
\caption{(Color Online.) Auto-spectrum of the polarization due to the primordial quadrupole at $z = 0.025$ (red, dot-dashed), $z = 0.425$ (green, dashed) and $z = 0.975$ (blue, solid).}
\label{fig:eta}
\end{figure}

\begin{figure}
\centering
\includegraphics[width=0.5\textwidth]{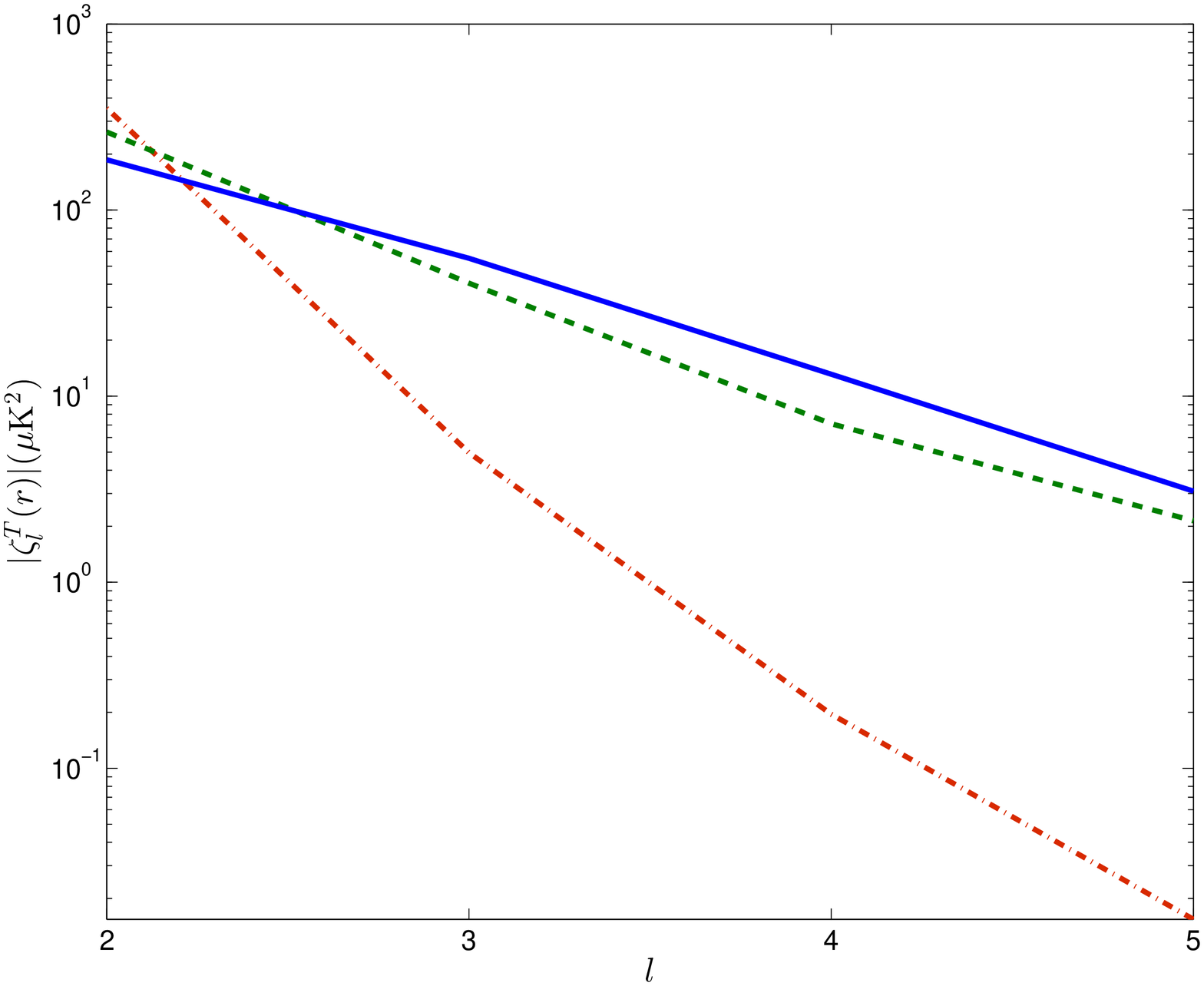}%
\includegraphics[width=0.5\textwidth]{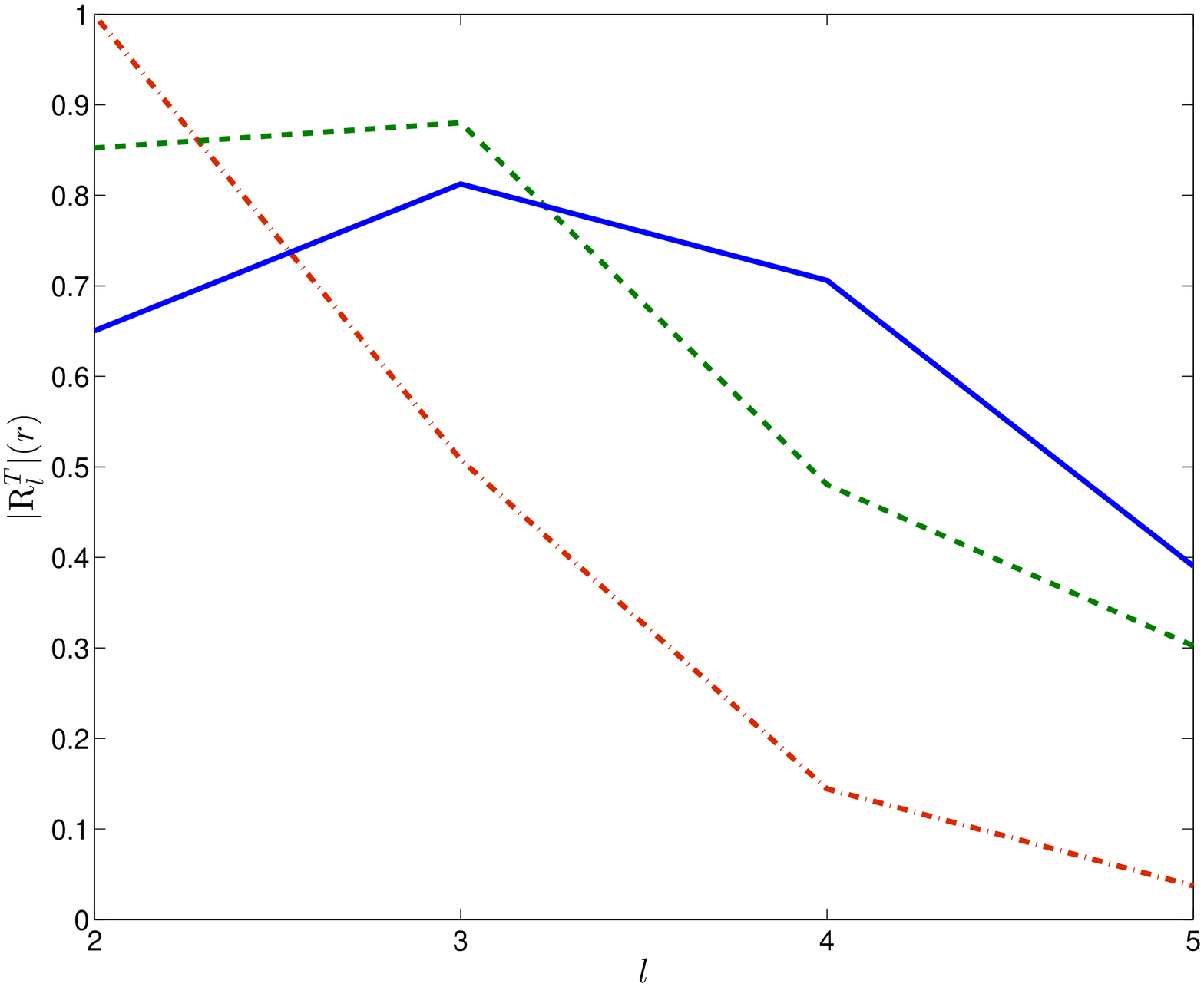}
\caption{(Color Online.) \textit{Left}: Cross-spectrum of the local CMB temperature and the cluster polarization due to the primordial quadrupole at $z = 0.025$ (red, dot-dashed), $z = 0.425$ (green, dashed) and $z = 0.975$ (blue, solid). \textit{Right}: Correlation coefficients of the cross-spectra on the left. Note that in our polarization convention, the cross-spectrum is negative over this range of $l$.}
\label{fig:zetaT}
\end{figure}

\begin{figure}
\centering
\includegraphics[width=0.5\textwidth]{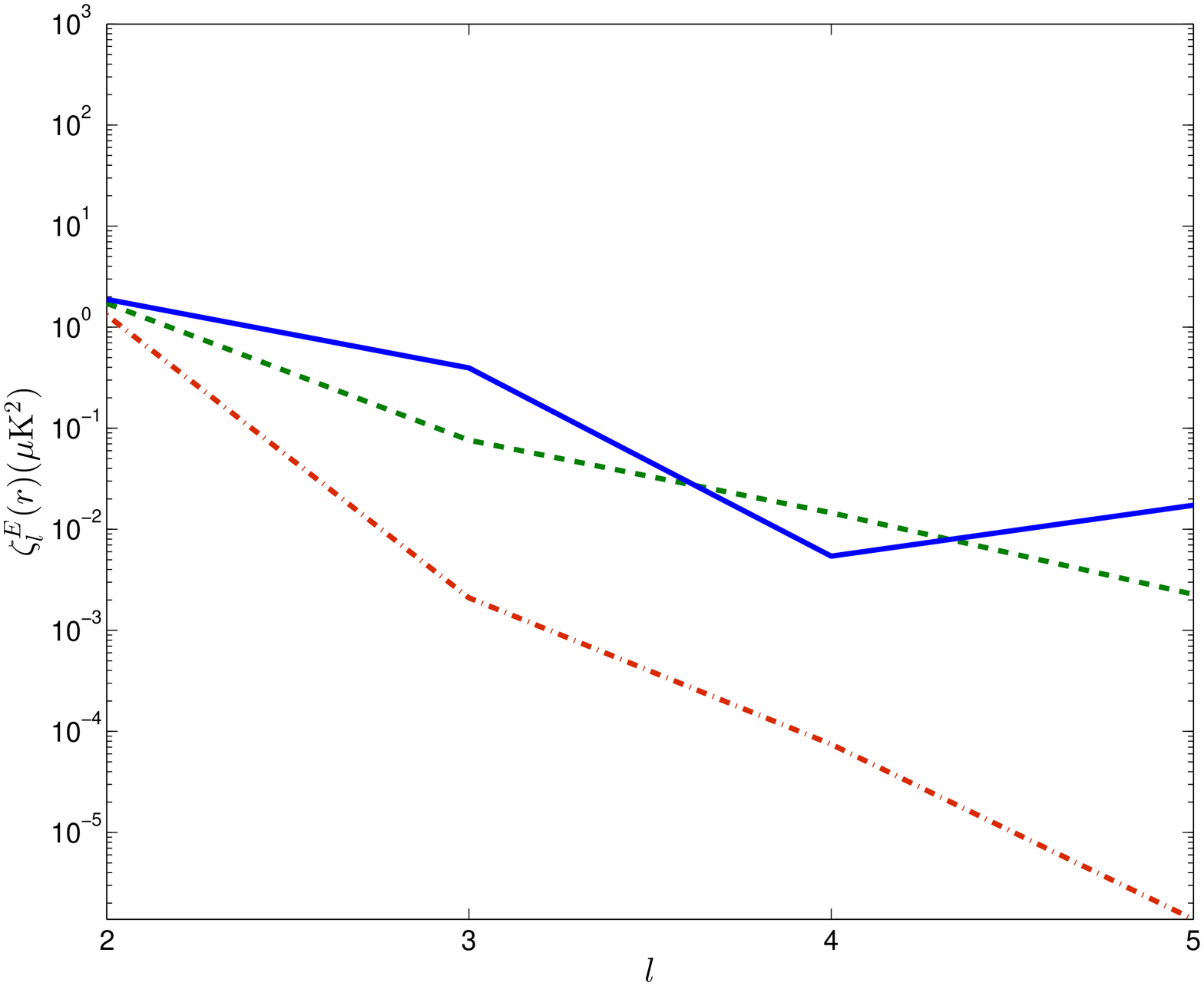}%
\includegraphics[width=0.5\textwidth]{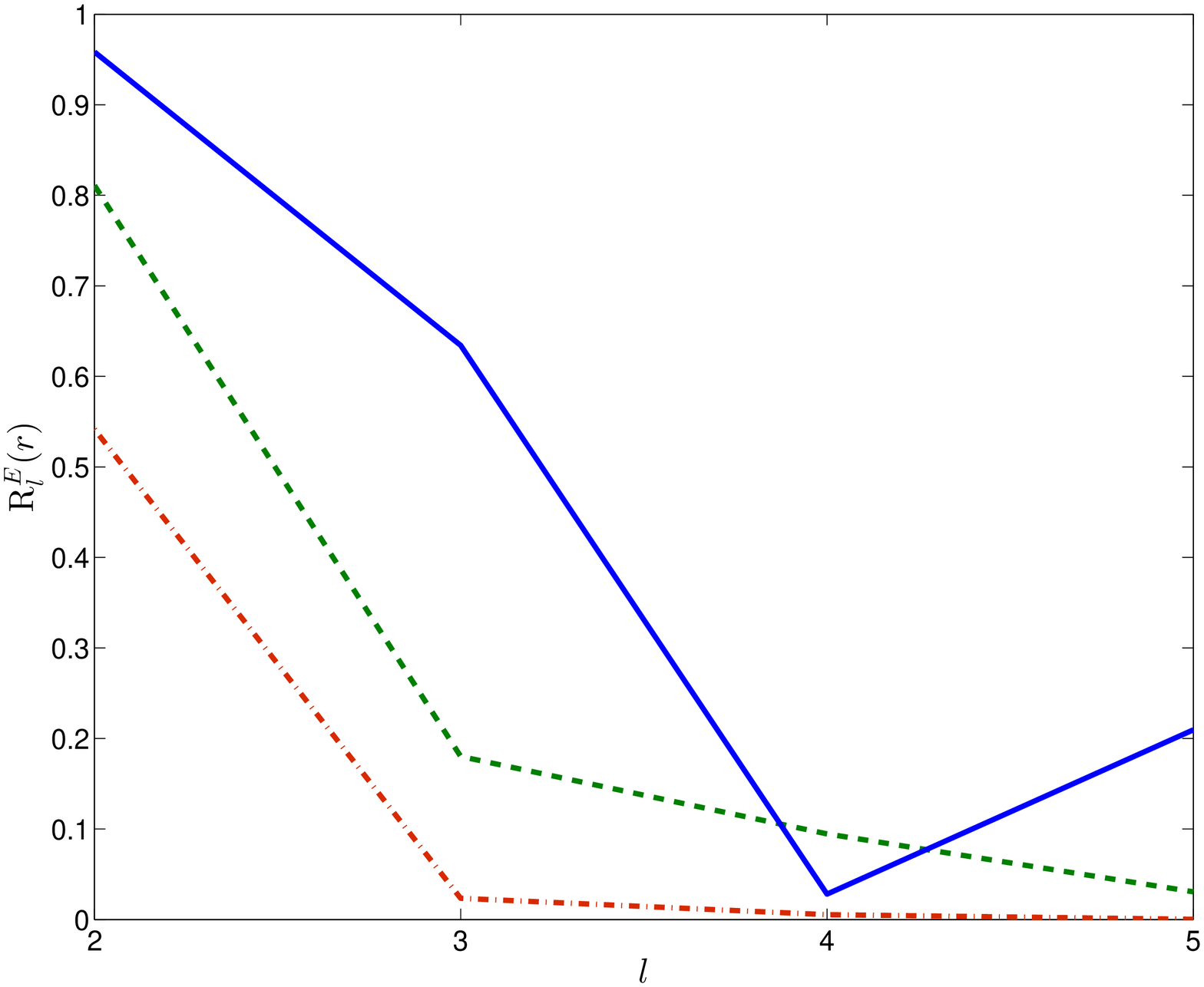}

\caption{(Color Online.) As Fig.~\ref{fig:zetaT}, but for the cross-spectrum of the local CMB $E$-mode and cluster polarization.}
\label{fig:zetaE}
\end{figure}

\begin{figure}
\centering
\includegraphics[width=0.6\textwidth]{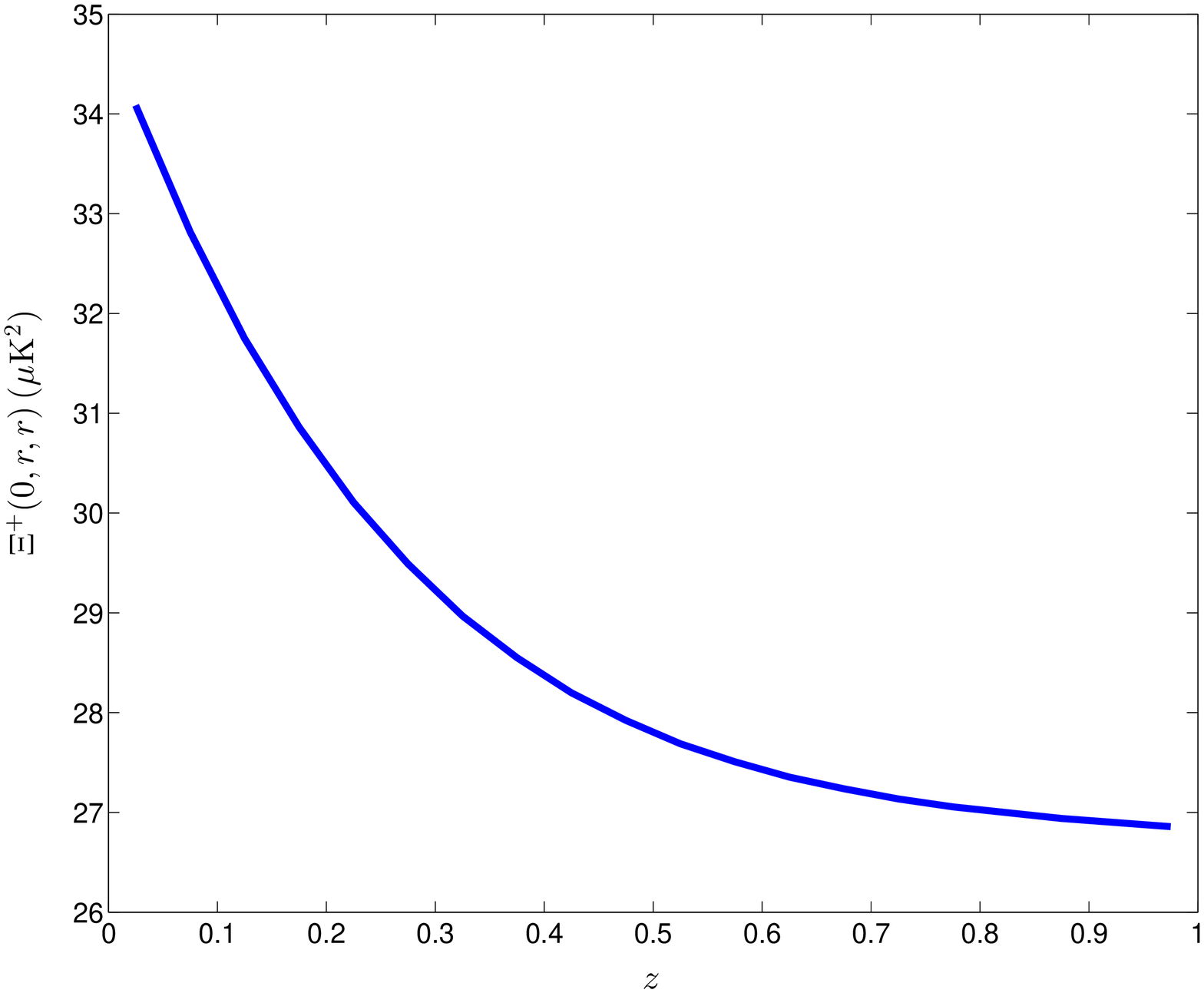}
\caption{(Color Online). Correlation function $\Xi^{+}$ at zero lag ($\beta = 0$) and equal redshift $z$.}
\label{fig:eta_zlag}
\end{figure}

In Fig.~\ref{fig:eta} we plot the auto-spectrum of the cluster polarization signal at three different redshifts, and in Figs.~\ref{fig:zetaT} and \ref{fig:zetaE} we plot the cross-spectrum with the local ($z=0$) CMB temperature and $E$-mode anisotropies. For plots of the cross-correlation between different redshifts, see \citep{2006PhRvD..73l3517B}. We also plot the correlation coefficient, defined as $R_l^X(r) = \zeta^X_l(r)/\sqrt{C_l^{XX} \xi_l(r,r)}$, where $X = T,E$. The dependence of the spectra in these plots on $l$ and $z$ is a combination of several effects. Firstly, at low redshift, the signal is almost entirely quadrupolar, with significant suppression of all the spectra at higher $l$. At higher redshift, the coherence scale of the quadrupole at the cluster redshift subtends smaller angular scales on the sky today, which boosts power in the higher multipoles, a consequence of free-streaming.  Secondly, the power in the quadrupole itself depends on redshift through the ISW term (and, weakly, through non scale-invariance of the primordial power spectrum). The onset of dark energy domination at $z \lesssim 1$ causes potentials to decay, sourcing the ISW effect. The quadrupole power is therefore larger at lower redshifts, which explains why $\xi_2(r,r)$ and $\zeta^T_2(r)$ decrease with increasing redshift. We can see this more clearly in Fig.~\ref{fig:eta_zlag}, where we plot the correlation function $\Xi^{+}$, defined in Eq.~\eqref{eq:corr_p}, at zero lag and equal redshift. Note that this is given by $\sum_l \frac{21+1}{4\pi}\xi_l(r,r)$, and hence is closely related to the integral of $l \xi_l(r,r)$, c.f. Fig.~\ref{fig:eta}. This quantity clearly rises at low redshift, due to the time-dependence of the quadrupole.

The cluster polarization signal also has a correlation with the local $E$-mode through $\zeta_l^E$. At large angular scales, the CMB $E$-mode is sourced by the scattering of the temperature quadrupole at reionization. The physics is exactly the same as that described above for galaxy clusters, except that the scattering probability (the visibility function of Thomson scattering) decays more slowly with redshift rather than being essentially a delta function in the cluster core. Since reionization occurs at high redshift ($z \sim 10$), the quadrupole coherence scale at reionization subtends smaller angular scales than at cluster redshifts. The greatest correlation will therefore be with polarization from high-redshift clusters. This behaviour can be clearly discerned in Fig.~\ref{fig:zetaE}, where the trend (for $l=2$--$5$) is for the correlation coefficient $R_l^E$ to be greater for clusters at higher redshift.

%This behaviour can be clearly discerned in Fig.~\ref{fig:zetaE}, where the correlation coefficient $R_l^E$ for high-redshift sources is larger at high $l$ relative to low $l$ than for low-redshift sources. However, the spectra still peak at $l=2$ due to the large contribution of the local temperature quadrupole for these low redshift sources.

\section{Likelihood Analysis and Optimal Estimator}
\label{sec:like}

In this section we construct the maximum-likelihood estimator for the cluster polarization signal, and discuss its statistics. We should begin by justifying the claim implicit in this section, viz.\ that a statistical treatment of the signal is necessary for detecting the quadrupole signal in forthcoming CMB polarization surveys.
The magnitude of the local temperature quadrupole is roughly $\sqrt{C^{TT}_2} \approx 16 \, \mu \K$ \citep{2013arXiv1303.5075P}, so a cluster at $z=0$ would source polarization $\sqrt{Q^2 + U^2}$ of order $2 \tau \, \mu \K$ in a pixel with optical depth $\tau$. Optical depths through the centres of clusters can be as high as $0.02$--$0.03$, so the polarization signal from these regions would be around $0.06 \, \mu \K$. A typical low-redshift cluster  has an angular radius (taken as $5R_{500}$ where $R_{\Delta}$ is the radius at which the enclosed density is $\Delta$ times the critical density at the cluster redshift) of roughly 10 arcmin. Ignoring the rapid fall-off of the optical depth with angular radius, and the effects of finite instrument resolution, a polarization survey with sensitivity of $1\,\mu\K$-$\rm{arcmin}$ for $Q$ and $U$ would give a signal-to-noise ($S/N$) around unity for a single cluster. Note that the polarization noise level of \emph{Planck} at 143~GHz for the full mission is expected to be around $60\,\mu\K$-$\rm{arcmin}$, and around $7\,\mu\K$-$\rm{arcmin}$ for the SPTPol survey at 150~GHz~\citep{2012SPIE.8452E..1EA}. We have also assumed that the cluster is at a `typical' point on the sphere, whereas in fact the magnitude of the polarization varies with angular position. As we shall see, even this estimate is optimistic due to the assumption of constant optical depth across the cluster.

% By comparison, the noise on this measurement ($Q \pm iU$) for a square pixel whose size is that of the \emph{Planck} 143 GHz beam is $\sqrt{2} \times 11.5 \, \mu \K$, using values from the \emph{Planck} blue book \citep{bluebook}. Neglecting correlations of the signal within a cluster, and accounting for the independent information brought by $Q$ and $U$, the signal-to-noise ($S/N$) from a single pixel would be around $0.06/11.5 \sim 0.005$. A typical low-redshift cluster  has an angular radius (taken as $5R_{500}$ where $R_{\Delta}$ is the radius at which the enclosed density is $\Delta$ times the critical density at the cluster redshift) of roughly 10 arcmin. The size of the $Planck$ beam at 143 GHz is around $7$ arcmin, so the cluster consists of roughly six independent pixels. Assuming the quadrupole is constant over the scale of the cluster, this improves the $S/N$ to about $0.01$. Note that the noise levels we have used are slightly larger than those of the current \emph{Planck} mission, and we have not accounted for the rapid fall of the optical depth as a function of angular radius. The optical depth should also be convolved with the beam function. We have also assumed that the cluster is at a `typical' point on the sphere, whereas in fact the magnitude of the polarization varies with angular position.

This back-of-the-envelope calculation tells us that the effect will not be detectable with forthcoming surveys in individual clusters, and so we will have to approach the problem statistically by stacking clusters appropriately. However, it is not immediately obvious as to how to do this optimally, as the signal is correlated over the sky, and we have to account for the relative orientations of the polarization basis at each point on the sphere. We will attempt to address these issues in the rest of this section.

\subsection{The conditional likelihood function}
\label{subsec:cond_like}

An obvious first step in the programme to use cluster polarization as a cosmological observable is to check that the signal is present with the expected amplitude. Our main objective in this paper is to assess the significance of this test with forthcoming surveys.
% Suppose one wanted to forecast the detectability of the polarization signal from scattering of the remote quadrupole at the locations of clusters in an SZ survey. How would one go about this? 
To this end, we parameterise the signal with an amplitude parameter $\alpha$, such that the observed polarization in a direction $\vnhat$ is modelled as
\begin{equation}
\label{eq:split0}
d(\vnhat) = \alpha \tau(\vnhat) p(\vnhat) + n(\vnhat),
\end{equation}
where $p(\vnhat)$ is given by Eq.~\eqref{eq:harm_exp} and $n(\vnhat)$ is a Gaussian instrumental noise contribution with zero mean and covariance $\mathbf{N}$. The parameter $\alpha$ takes a fiducial value of unity. Estimating $\alpha$ then requires us to write down a likelihood for the data. But since we do not know the values of $p(\vnhat)$, we will have to marginalise over it. The resultant distribution will be Gaussian with a variance which is the sum of the noise variance and the cosmic variance given by $\xi_l(r,r')$.

However this cosmic variance is potentially very large, since the signal is on large scales. The variance on the measured $\alpha$ is correspondingly large and the test of concordance with the expected value $\alpha=1$ is weak. The situation is analogous to that of the large-scale CMB. We can make very accurate measurements of our temperature quadrupole, and yet our measurement of the \emph{true} (ensemble-averaged) quadrupole, given by $C^{TT}_2$, is very poor due to cosmic variance.

We can make a more powerful test by exploiting the correlation between the cluster signal and the local large-angle CMB anisotropies. Where the correlation is weak, we have to use the power observed in $p(\vnhat)$ to constrain $\alpha$. However, at low redshift most of the signal is correlated with the local temperature quadrupole and the first few higher-order moments (see Fig.~\ref{fig:zetaT}), and so the expected signal can be accurately predicted. By cross-correlating the measured cluster signal with the large-angle local anisotropies, our test of $\alpha=1$ will be limited only by noise in the cluster measurement and the fluctuations in the uncorrelated component of the cluster signal.

% Suppose all the clusters in our SZ survey were at $z=0$. The signal would then be given by Eq.~\eqref{eq:pol_in}, i.e. a discrete set of samples of the projected local temperature quadrupole modulated by the optical depth profile of each cluster. With prior knowledge of the optical depths, the signal is maximally correlated with the local quadrupole, which has been measured with high accuracy by space-based CMB experiments. This strongly suggests that any attempt to measure the amplitude of the remote quadrupole signal should be made conditional on the local CMB temperature (and, in principle, polarization) anisotropies.

% We still have no prior knowledge of the part of the signal not correlated with the local CMB, and so this part should still be marginalized over. However, this term is only important for high-redshift clusters, so for the surveys we consider the signal is essentially fixed with the experimental noise the dominant source of variance. By setting up the problem in this way, our forecasts faithfully represent the ability of an experiment to measure this signal. 

We split the observed polarization as
\begin{equation}
\label{eq:split}
d(\vnhat) = \alpha \tau(\vnhat)[p_c(\vnhat) + p_u(\vnhat)]+ n(\vnhat),
\end{equation}
where the part of the signal correlated with the local temperature is 
\begin{equation}
\label{eq:p_c}
p_c(\vnhat) =  \sum_{lm}\frac{\zeta_l^T(z)}{C_l^{TT}}a_{lm}{}_{-2}Y_{lm}(\vnhat),
\end{equation}
(we ignore correlation with local large-angle polarization, although this could be included straightforwardly) and the uncorrelated part is given by
\begin{equation}
p_u(\vnhat) = p(\vnhat) - p_c(\vnhat).
\end{equation}
The likelihood conditional on the large-angle temperature anisotropies (ignoring noise in their measurement) is, up to an $\alpha$-independent constant,
\begin{equation}
\label{eq:loglikefull}
-2\ln{L(d | \alpha, p_c)} = (\mathbf{d} - \alpha \mathbf{P}_c)^{\dag}(\mathbf{N} + \alpha^2 \mathbf{C}_u)^{-1}(\mathbf{d} - \alpha \mathbf{P}_c) + \ln \det(\mathbf{N} + \alpha^2 \mathbf{C}_u) ,
\end{equation}
where $\mathbf{d} \equiv d(\vnhat)$, and we have defined $\mathbf{P} \equiv \tau(\vnhat)p(\vnhat)$. The covariance matrix of $\mathbf{P}_u$ is $\mathbf{C}_u$. Note that these covariance matrices are defined with respect to the local $Q/U$ basis at each cluster, and so each element requires the appropriate rotation factors of Eqs.~\eqref{eq:p_rot1} and \eqref{eq:p_rot2} to be related to the physically-defined correlation functions of Eqs.~\eqref{eq:xi}--\eqref{eq:zetaE}.

We shall derive an approximate maximum-likelihood estimator for $\alpha$ based on the conditional likelihood of Eq.~\eqref{eq:loglikefull} in Sec.~\ref{sec:hi-z}. The $\alpha^2$ factors multiplying $\mathbf{C}_u$ complicate the construction of a maximum-likelihood estimator. However, this dependence of the likelihood on $\alpha$ is only important for high-redshift clusters --- it represents the additional information that can be obtained on the amplitude $\alpha$ from the power in $p_u(\vnhat)$. For the near-future surveys we consider, instrument noise and the moderately low-redshift cluster samples mean that we can ignore the $\alpha$ dependence of the likelihood through the sub-dominant $\mathbf{C}_u$, leaving
\begin{equation}
\label{eq:loglike}
-2\ln{L(d | \alpha, p_c)} \approx (\mathbf{d} - \alpha \mathbf{P}_c)^{\dag}(\mathbf{N} + \mathbf{C}_u)^{-1}(\mathbf{d} - \alpha \mathbf{P}_c),
\end{equation}
up to an additive constant. The maximum-likelihood estimate of $\alpha$ (see below) is now linear in the data and is an optimally-weighted measure of the cross-correlation between the cluster polarization and the large-angle temperature anisotropies across angular scales and redshifts.

We shall assume that the noise contribution to each pixel is constant and uncorrelated between pixels, such that $\mathbf{N}$ has diagonal elements given by $\sigma_N^2 = 2 \sigma_P^2$, with $\sigma_P^2$ the variance on the measured $Q$ (or $U$). The vector $\mathbf{d}$ has length $2 \times N_{c} \times \overline{N}_{\mathrm{pix}}$, containing the measured $Q+iU$ and $Q-iU$ values in each pixel, with the $\overline{N}_{\mathrm{pix}}$ the average number of pixels in a cluster, and $N_c$ the total number of clusters.
\subsection{Demodulating the optical depths}
\label{subsec:opticaldepth}

The correlation length of the CMB quadrupole is much larger than the extent of individual clusters, and so the quadrupole may be treated as constant across the cluster. [This assumption was already made in Eq.~\eqref{eq:harm_exp}.] The observed polarization is therefore the signal of interest multiplied by the beam-convolved optical depth. For each cluster we can compress the measurements into a single estimate of the polarization for that cluster $\hat{p}^{(j)}$ (where bracketed indices here refer to clusters):
% It will prove useful at this stage to remove the dependence of the signal on the beam-convolved optical depth profile of each cluster. The key property of the signal that allows us to do this is the fact that the correlation length of the quadrupole is much greater than the size of a typical cluster, such that the quadrupole can be approximated as constant in pixels within a given cluster. We can then rewrite the likelihood (conditional on the \emph{total} polarization) in the following form
% %
% \begin{equation}
% \label{eq:loglike2}
% -2\ln{L(d | p)} = \sum_{i,j,k,m}(d_i^{(j)} - \tau_i^{(j)}p^{(j)})^{*}(\mathbf{N})^{-1}_{i,(j),k,(m)}(d_k^{(m)} - \tau_k^{(m)} p^{(m)}),
% \end{equation}
% %
% where bracketed indices refer to clusters and unbracketed indices refer to pixels within a cluster. 
% The maximum-likelihood estimate of $p^{(j)}$ is then
%
\begin{equation}
\label{eq:phat}
\hat{p}^{(j)} = \frac{\bm{\tau}_{(j)}^{\top}\mathbf{N}^{-1}\mathbf{d}^{(j)}}{\bm{\tau}_{(j)}^{\top}\mathbf{N}^{-1}\bm{\tau}_{(j)}} = \frac{\sum_i \tau_i^{(j)} d_i}{\sum_i \left(\tau_i^{(j)}\right)^2},
\end{equation}
where, temporarily, bold-faced notation refers to pixels within a cluster, and $\bm{\tau}_{(j)}$ is the beam-convolved optical depth for cluster $j$. There is no sum over $j$ implied.
The variance on this estimator is
\begin{equation}
\label{eq:phat_var}
\mathrm{var}(\hat{p}^{(j)}) = (\bm{\tau}_{(j)}^{\top}\mathbf{N}^{-1}\bm{\tau}_{(j)})^{-1} = \frac{\sigma_N^2}{\sum_i \left(\tau_i^{(j)}\right)^2} .
\end{equation}

The approximate likelihood~\eqref{eq:loglike} becomes for the demodulated signal
\begin{equation}
\label{eq:loglike3}
-2\ln{L(\hat{p} | \alpha, p_c)} = (\hat{\mathbf{p}} - \alpha \mathbf{p}_c)^{\dag}(\mathbf{C}_{\hat{p}} + \mathbf{C}_u)^{-1}(\hat{\mathbf{p}} - \alpha \mathbf{p}_c),
\end{equation}
where bold-faced notation now refers to clusters, and the thermal noise variance now includes the optical depth weighting $(\mathbf{C}_{\hat{p}})_{ij} = \mathrm{var}(\hat{p}^{(j)})\delta_{ij}$. This procedure has significantly reduced the dimensionality of the matrix inversions; the distribution in Eq.~\eqref{eq:loglike3} now has $2 \times N_{c}$ degrees of freedom.

Finally we note that the above demodulation can also be done in harmonic space, which should be very accurate for a survey with full-sky coverage. We outline this procedure in Appendix~\ref{app:harm}.

\subsection{Maximum-likelihood estimator}
\label{subsec:MLestimator}

Using Bayes' theorem and assuming a flat prior on the parameter $\alpha$, the likelihood of Eq.~\eqref{eq:loglike3} is proportional to the posterior distribution of $\alpha$ given the data. We may then maximize this distribution with respect to $\alpha$ to find the maximum-likelihood estimator
\begin{equation}
\label{eq:alpha_hat}
\hat{\alpha} = \frac{\operatorname{Re} [\mathbf{p}_c^{\dag}(\mathbf{C}_{\hat{p}} + \mathbf{C}_u)^{-1}\hat{\mathbf{p}}]}{\mathbf{p}_c^{\dag}(\mathbf{C}_{\hat{p}} + \mathbf{C}_u)^{-1}\mathbf{p}_c},
\end{equation}
which has variance
\begin{equation}
\label{eq:sn}
\mathrm{var}(\hat{\alpha}) = \left(\mathbf{p}_c^{\dag}(\mathbf{C}_{\hat{p}} + \mathbf{C}_u)^{-1}\mathbf{p}_c\right)^{-1}.
\end{equation}
Reassuringly, $\hat{\alpha}$ and its variance are independent of the polarization basis used to define $Q$ and $U$. To see this, note that a change of basis is effected by a unitary (here diagonal) transformation that cancels in forming the scalar $\hat{\alpha}$.

The estimator, Eq.~\eqref{eq:alpha_hat}, has a simple physical interpretation. In the limit that $\mathbf{C}_u \rightarrow 0$, the numerator of Eq.~\eqref{eq:alpha_hat} takes the form of a correlation between the inverse-variance weighted measured polarization $(\mathbf{C}_{\hat{p}})^{-1}\hat{\mathbf{p}}$ and $\mathbf{p}_c$. The latter is the predicted signal given measurements of our local temperature field and knowledge of its correlation with the cluster polarization. This prediction should be precisely the underlying signal in the $\mathbf{C}_u \rightarrow 0$ limit, so the best estimate of the signal amplitude is just the inverse-variance-weighted correlation, with a normalisation given by the denominator of Eq.~\eqref{eq:alpha_hat}. For clusters at higher redshift, we must include the cosmic variance of the uncorrelated part of the signal, and compare this inverse-variance-weighted polarization with the local prediction.

In formulating the estimator, Eq.~\eqref{eq:alpha_hat}, we have ignored the fact that we are not randomly sampling the polarization field $p(\vnhat;r)$, since clusters are biased tracers of large-scale structure. Essentially, our observable is the polarization field weighted by the cluster number density (and survey selection function), and complications arise from the correlations between the weight function and the local temperature anisotropies and the cluster polarization field $p(\vnhat;r)$. For example, the average of Eq.~\eqref{eq:alpha_hat} conditioned on the local temperature anisotropies \emph{and} the cluster number density in our universe is no longer unity. This is because the (doubly) conditional average of $p(\vnhat;r)$ is not simply $p_c(\vnhat;r)$, but has an additional term from the part of $p(\vnhat;r)$ that is uncorrelated with the local temperature anisotropies but is correlated with the cluster number density. However, such a bias is expected to be small for low-redshift cluster surveys since very little of $p(\vnhat;r)$ is uncorrelated with the local temperature anisotropies. Furthermore, since we are dealing with very large-scale fields, it is only the cluster number density smoothed on scales approaching those of the cluster survey itself that are relevant, further reducing its impact. A detailed analysis of these issues may potentially be important for higher redshift surveys, and would best be addressed through simulations.

\section{Signal-to-noise forecasts}
\label{sec:sn}

In this section we forecast the $S/N$ for the estimator in Eq.~\eqref{eq:alpha_hat} for present and future SZ polarization surveys. We begin by detailing our assumptions about these surveys and the cluster sample they expect to observe.

% Firstly we note that the optical depths that appear in Eqs.~\eqref{eq:phat} and \eqref{eq:phat_var} are actually the \emph{beam-convolved} optical depths measured by the telescope. In our forecasts, we explicitly perform this convolution, assuming a Gaussian beam function with a full-width-at-half-maximum (FWHM) given by $\theta_{\mathrm{fwhm}}$.

\subsection{Cluster sample assumptions}

To forecast future constraints, we need to assume optical depth profiles for each cluster in the survey. The optical depths are defined as integrals of the free electron density through each cluster. Assuming complete ionization in the ICM, this is equivalent to an integral over the gas density $\rho_g$. 

We consider two different phenomenological models for the gas density profile: the common $\beta$-model parameterisation, and a more realistic cored Navarro-Frenk-White (cNFW) profile.

\subsubsection{$\beta$-model}

In the isothermal $\beta$-model, the gas density profile is spherically symmetric with a radial profile
\begin{equation}
\label{eq:beta}
\rho_g(r;z) = \rho_g^0(z)\left(1+\frac{r^2}{r_s^2}\right)^{-3\beta/2},
\end{equation}
where $\rho_g^0(z)$ is a (redshift-dependent) normalization factor. In keeping with a conventional choice, we take $\beta=2/3$. 

To fix the normalization, we assume that the ratio of gas mass to total mass within the virial radius (taken to be $R_{200}$) is equal to the cosmological value. To calculate the dark matter mass within the virial radius, we assume an NFW profile
\begin{equation}
\label{eq:NFW}
\rho_m(r;z) = \frac{\rho_m^0(z)}{(r/r_s)(1+r/r_s)^2}.
\end{equation}
The normalization $\rho_m^0$ can be expressed in terms of the critical density at the cluster redshift and the concentration parameter $c_{500}$, which is related to the scale radius $r_s$ by $c_{500} = R_{500}/r_s$. We fix $c_{500} = 1.177$, which is the best-fit value for the universal pressure profile of \citep{2010A&A...517A..92A}, derived from X-ray observations. This model thus has one free parameter, given by $R_{500}$.

\subsubsection{Cored NFW model}

The $\beta$-model is known to be a rather poor description of the gas density in a typical cluster, particularly in the outer regions \citep{2006ApJ...640..691V}. Instead, a combination of X-ray measurements and hydrodynamic $N$-body simulations suggests that the gas density falls more steeply than Eq.~\eqref{eq:beta} at large radii, tracing the underlying dark matter halo quite faithfully for high-mass systems \citep{2010A&A...517A..92A, 2007MNRAS.380..877S}. Motivated by this, we consider another model taken to be an NFW profile with a core, given by
\begin{equation}
\label{eq:cNFW}
\rho_g(r;z) = \frac{\rho_g^0(z)}{[(r+r_0)/r_s](1+r/r_s)^2},
\end{equation}
where $r_0$ is the core radius. We assume that $r_0/R_{200} = 0.02$ is fixed for all clusters as in \citep{2006MNRAS.366..397S}, although this choice does not affect our results significantly. We again fix $r_s$ through $c_{500} = R_{500}/r_s = 1.177$, leaving $R_{500}$ as a free parameter, and fix the normalization in the same way as for the $\beta$-model.

\subsubsection{Optical depth profiles}

The optical depth through the cluster at angular radius $\theta$ from the centre is given by
\begin{equation}
\label{eq:optdepth}
\tau(\theta;z) = \sigma_{T} \int^{L}_{-L} n_e(r(l);z) \mathrm{d} l,
\end{equation}
where $r(l) = \sqrt{l^2 + (d_A \theta)^2}$, $L$ is a cut-off (in principle $\theta$-dependent, see below), and $d_A$ is the angular diameter distance. The free electron density is related to the gas density through $n_e(r;z) = (1+f_H)\rho_g(r;z)/(2 m_p)$ where $f_H$ is the hydrogen mass fraction (taken to be 0.76) and $m_p$ is the mass of the proton.

The simple analytic gas profiles we have considered are not expected to hold in the cluster outskirts, particularly for low-mass systems ($M \lesssim 10^{14} M_{\odot}$) where accretion could be important. We therefore choose to truncate the profiles at $5R_{500}$, which corresponds to roughly 3.1 virial radii. This may be an optimistic choice given the current deficit of observations beyond the virial radius, but it should not affect the results too much due to the low gas density in the outskirts. 

When computing the integral in Eq.~\eqref{eq:optdepth}, there is some freedom as to the choice of cut-off $L$. Given the uncertainty in the gas models at large physical cluster radii, large angular radii can be excluded from the analysis, but we will always have to deal with the unmodelled region along the line of sight. Simply excluding this region by truncating the integral at radii larger than, say, $5R_{500}$ will underestimate the optical depth at all angular positions across the cluster. In this work, we set $L$ large enough so that all line-of-sight gas is included in the optical depth calculation. Other reasonable choices of cut-off $L$ lead to roughly $10\%$ variations in our final $S/N$ estimates.

With the above assumptions, typical central optical depths for our cluster sample are 0.01 and 0.004 at $z=0$ and 0.04 and 0.01 at $z=1$ for the cNFW and $\beta$ models respectively. The $\beta$-model predicts a higher gas density in the outskirts, but this is more than compensated for by the higher central density in the cNFW model, which explains the differences in central optical depths. Note that the normalization of both models implies $\tau \propto R_{500} \rho_{\mathrm{crit}}(z)$ and hence clusters of a given physical size have higher optical depths at higher redshifts, having formed in denser environments.

\begin{figure}
\centering
\includegraphics[width=0.6\textwidth]{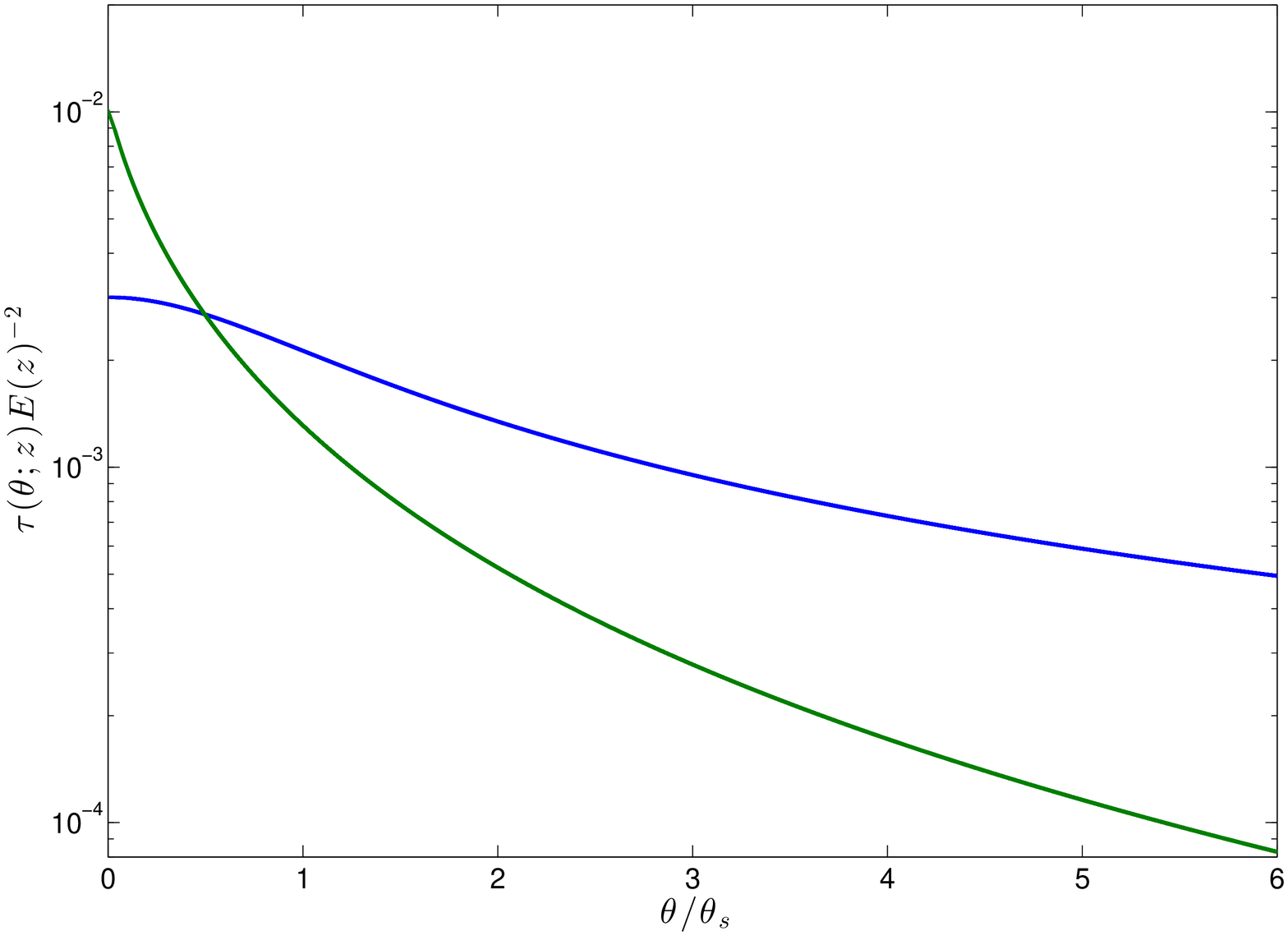}
\caption{(Color Online.) Scaled optical depth profiles as a function of the scaled angular radius $\theta/\theta_s$, where $\theta_s = r_s/d_A$, for the $\beta$-model (blue, top solid), and the cNFW model (green, lower solid). We have assumed a cluster with $R_{500} = 1 \, \mathrm{Mpc}$. Note that the optical depth profiles used in the $S/N$ calculation are truncated at $\theta/\theta_s = 5c_{500} = 5.885$.}
\label{fig:tau}
\end{figure}

In Fig.~\ref{fig:tau} we plot the redshift-independent quantity $\tau(\theta;z) E(z)^{-2}$, where $E(z) = H(z)/H_0$ is the dimensionless Hubble parameter, assuming $R_{500} = 1 \, \mathrm{Mpc}$.

Finally, we note that it is the beam-convolved optical depth that is relevant for observations of cluster polarization. In our forecasts, we explicitly perform this convolution, assuming a Gaussian beam function with a full-width-at-half-maximum (FWHM) given by $\theta_{\mathrm{fwhm}}$.

\subsection{Survey assumptions}

We forecasts constraints on $\alpha$ for a full-sky mission similar to \emph{Planck} and for the 150\,GHz channel of the South Pole Telescope polarization experiment (SPTPol; \citep{2012SPIE.8452E..1EA}). For the \emph{Planck}-like survey, we use a single frequency of 143\,GHz with specifications taken from the  \emph{Planck} blue book: thermal noise on $Q$ (or $U$) of 11.5 $\mu \K$ in a square pixel of side length $\theta_{\mathrm{fwhm}} = 7.1$\,arcmin. (We would expect that combining other frequency channels on \emph{Planck} would increase our $S/N$ forecasts by less than $\sqrt{2}$.) For SPTPol, the 150\,GHz channel was chosen for its superior sensitivity and resolution over the 90\,GHz channel. We take specifications from the SPTPol overview paper \citep{2012SPIE.8452E..1EA}: polarization noise of $7.1 \, \mu \K$ for a $\theta_{\mathrm{fwhm}} = 1.0$ arcmin pixel.

The Atacama Cosmology Telescope is also performing polarization surveys~\citep{2010SPIE.7741E..51N,2014arXiv1405.5524N}. The final forecast sensitivities are similar to SPTPol, so our forecasts for SPTPol should be fairly similar, up to the different field positions that we discuss later.

\begin{figure}
\centering
\includegraphics[width=0.6\textwidth]{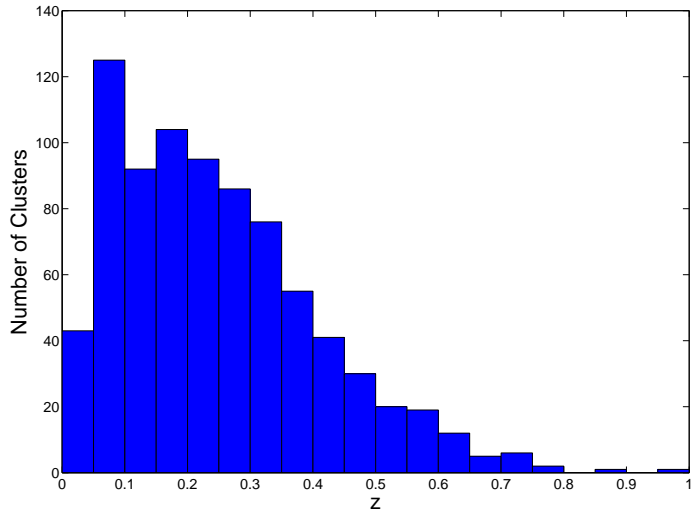}
\caption{(Color Online.) Redshift distribution of SZ clusters from the \emph{Planck} 2013 catalogue.}
\label{fig:planck_bins}
\end{figure}

For our cluster distribution, we assume that all the clusters used in the polarization survey have already been discovered or will soon be discovered by the SZ surveys of \emph{Planck} and SPT. For \emph{Planck}, we take the redshift distribution of the 813 clusters with confirmed redshifts from the \emph{Planck} 2013 SZ catalogue \citep{2013arXiv1303.5089P}, which consists of 20 redshift bins equally spaced between central values of 0.025 and 0.975. This distribution is plotted in Fig.~\ref{fig:planck_bins}. We then use the X-ray angular sizes of clusters in the smaller \emph{Planck} Early SZ catalogue (ESZ; \citep{2011A&A...536A...8P}) to deduce typical angular sizes in the redshift bins of the full distribution, by re-binning the ESZ samples assuming constant sizes in each bin. This translates to a roughly constant physical cluster size of $R_{500} \approx 1 \, \mathrm{Mpc}$ in each bin. In the lowest redshift bin, we use the 1-$\sigma$ lower limit of the mean size in the re-binned ESZ catalogue, to avoid including artificially large low-redshift clusters that were not present in the ESZ sample. Note that the approximate constancy of physical cluster radii is a statement about the selection function of the survey rather than the whole cluster population, for which the sizes are definitely not constant at different redshifts. 

\begin{figure}
\centering
\includegraphics[width=0.6\textwidth]{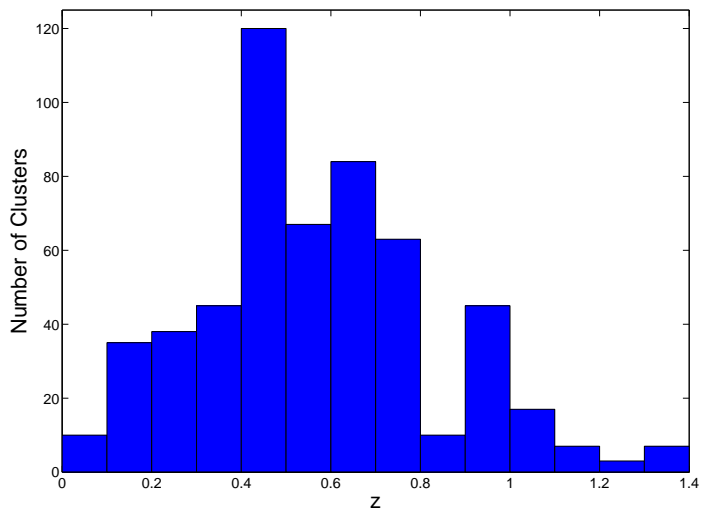}
\caption{(Color Online.) Forecast redshift distribution of SZ clusters for SPTPol.}
\label{fig:spt_bins}
\end{figure}

For the SPTPol survey, we take the redshift distribution of \citep{2013ApJ...763..127R} supplemented with an extra 400 clusters expected from the full SPT survey of 2500 $\mathrm{deg}^2$ \citep{2013ApJ...763..127R}. This amounts to 558 clusters in total, distributed in 14 redshift bins plotted in Fig.~\ref{fig:spt_bins}. We assume a constant cluster size of $R_{500} = 1\,\mathrm{Mpc}$ to calculate angular sizes in each bin\footnote{For SPT, the assumption of constant $R_{500}$ should be reasonable, as suggested by the following argument. Given the high resolution, the sample selection should roughly correspond to a minimum integrated Compton $y$ (i.e., $Y = \int d^2\vnhat\, y $). With the cluster assumptions here, $Y \propto H^2(z) R_{500}^3 / d_A^2(z)$ for a cluster of given $R_{500}$ at redshift $z$. The ratio $H^2/d_A^2$ is very flat around its minimum at $z\approx 0.7$ so, over the SPT redshift range, constant $Y$ corresponds to roughly constant $R_{500}$, and in any redshift bin the survey should contain all clusters larger than a given redshift-independent $R_{\rm min}$. Since $d n / dR_{500} \propto M^{-1/3} dn / d\ln M$, where $dn  / d\ln M$ is the cluster mass function, the most numerous clusters in any bin will be those with radius close to $R_{\rm min}$. Exactly how ``most numerous'' relates to the mean radius depends on where the limiting radius (hence mass) lies on the mass function at the given redshift. The approximation should be more accurate if the cut is on the steeper (i.e., high-mass) part of the mass function.}.

Typical noise variances, defined by Eq.~\eqref{eq:phat_var}, are then $(2.09 \,\mathrm{mK})^2$ for the \emph{Planck}-like survey at $z=0.025$ in the cNFW model, rising to $(12.23 \,\mathrm{mK})^2$ at $z=0.975$. In the $\beta$-model, the corresponding values are $(1.17 \,\mathrm{mK})^2$ and $(4.76 \,\mathrm{mK})^2$ respectively. For SPTPol, in the cNFW model the noise variance is $(111.7 \, \mu \mathrm{K})^2$ at $z=0.05$ and $(260.7 \, \mu \mathrm{K})^2$ at $z=1.35$. In the $\beta$-model the corresponding values are $(66.1 \, \mu \mathrm{K})^2$ and $(142.5 \, \mu \mathrm{K})^2$.

We position clusters on the sky at random in each redshift bin, assuming full sky coverage, $f_{\mathrm{sky}} = 1$, for the \emph{Planck}-like survey, and a 625 $\mathrm{deg}^2$ field centred at $(23^{\mathrm{h}}30^{\mathrm{m}}, -55^{\circ})$ in celestial coordinates for SPTPol \citep{2012SPIE.8452E..1EA}. Although the SPTPol survey covers less area than the full SPT-SZ survey, the number of expected clusters is roughly the same, since the mass threshold of SPTPol is lower \citep{2012SPIE.8452E..1EA}.

\begin{figure}
\centering
\includegraphics[width=0.5\textwidth]{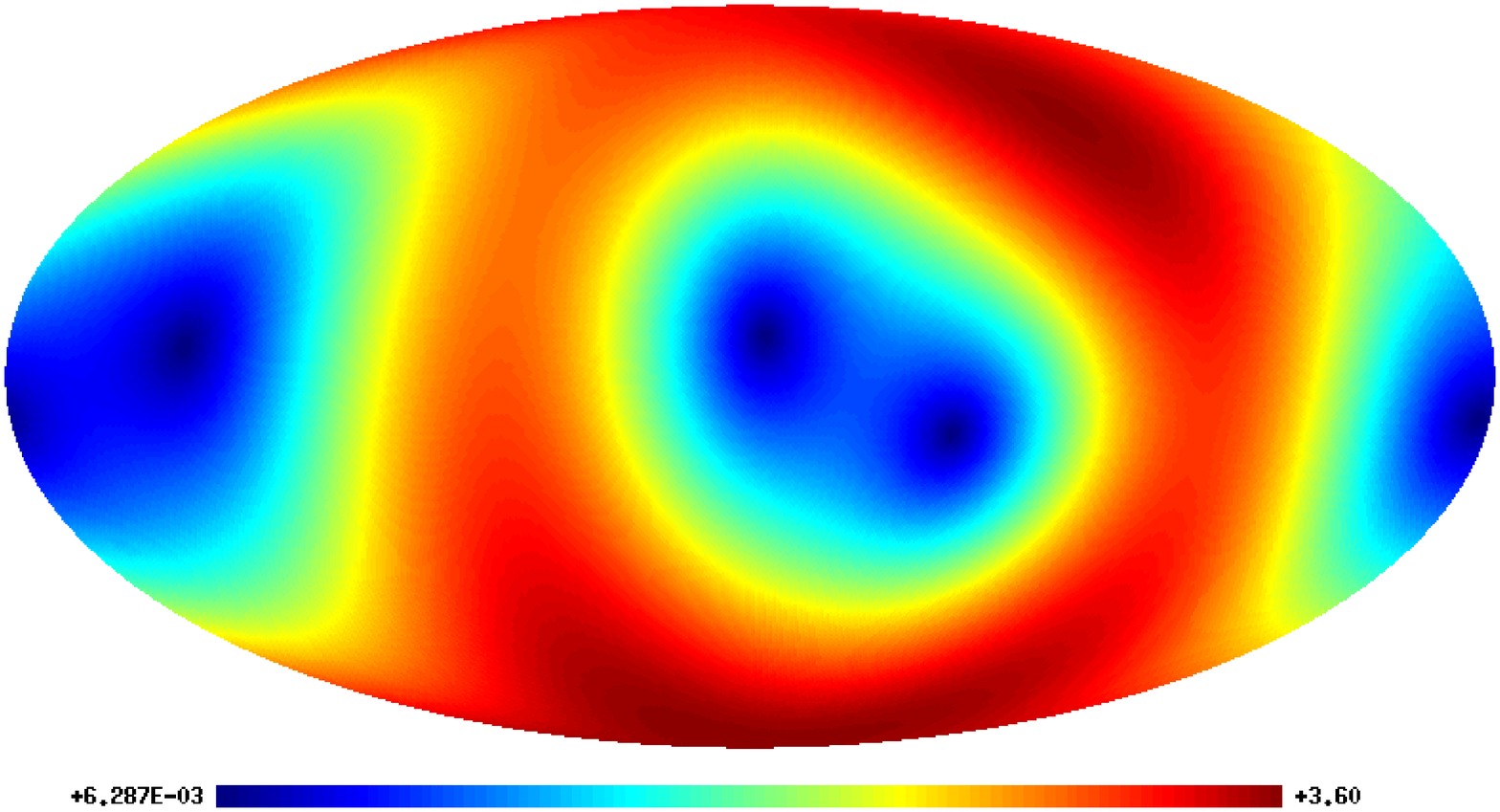}
\includegraphics[width=0.5\textwidth]{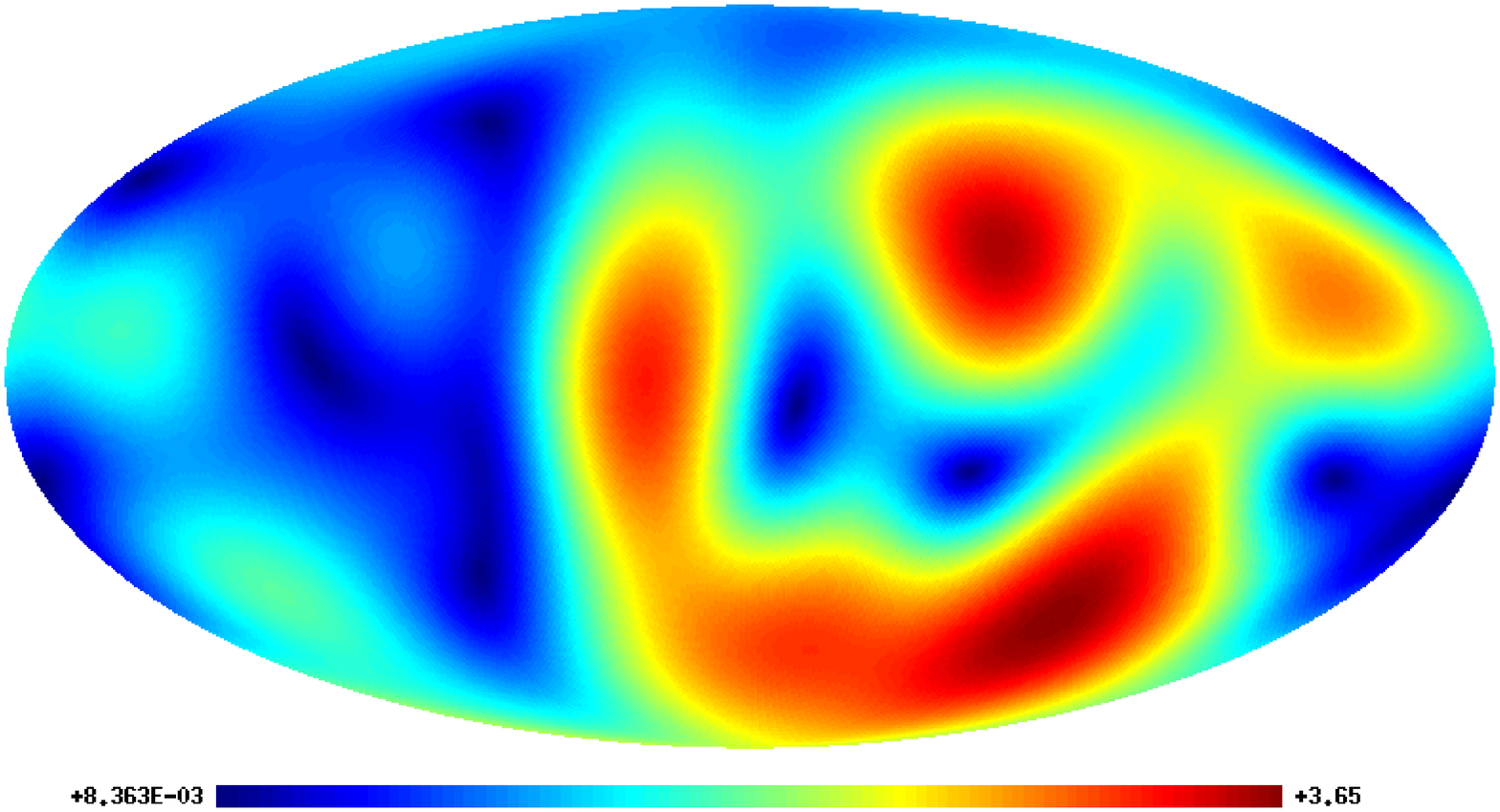}
\caption{(Color Online.) Magnitude of the correlated part of the polarization field sourced at $z=0.05$ (top) and $z=1.35$ (bottom) in Galactic coordinates. Units are $\mu \mathrm{K}$.}
\label{fig:maps}
\end{figure}

Finally, we extract the first few temperature multipole coefficients from the 2013 \emph{Planck} SMICA map \citep{2013arXiv1303.5072P}, using the \textsc{HEALPix}\footnote{http://healpix.sourceforge.net} package, for use in calculating the correlated polarization from Eq.~\eqref{eq:p_c}. We find that only the $l\leq 5$ are required for convergence in our signal-to-noise calculations. In Fig.~\ref{fig:maps} we show the resultant correlated polarization magnitude $\sqrt{Q_c^2 + U_c^2}$ for $z=0.05$  and $z=1.35$. As discussed in Sec.~\ref{sec:physics}, higher redshift sources have greater angular structure when observed today due to free-streaming.

\subsection{Signal-to-noise forecasts}

After convolving the optical depth profile of each cluster with the telescope beam profile, we can evaluate the $S/N$ on $\alpha$ from Eq.~\eqref{eq:sn}. Note that the $\tau$-weighted noise variance has a simple form in Fourier space due to Parseval's theorem, allowing for rapid computation of $\mathbf{C}_{\hat{p}}$.

For the \emph{Planck}-like survey, we forecast a $S/N$ on $\alpha$ of 0.03 for the cNFW model, and 0.06 for the $\beta$-model. The higher value for the $\beta$-model is due to the higher gas density in the cluster outskirts carried by this model, which carry more weight in the calculation of $\mathbf{C}_{\hat{p}}$ due to these regions making up most of the angular extent. By comparison, the higher central density of the cNFW profile is diluted somewhat by the poor resolution of \emph{Planck}.

For SPTPol, we forecast a $S/N$ of 0.28 for the cNFW model and 0.48 for the $\beta$-model. The superior resolution of SPT allows much more independent information to be extracted from the clusters, and is particularly helpful for highly concentrated clusters since this is where the signal is greatest.

Note that in both experiments, the cluster redshifts are sufficiently low that the noise variance term $\mathbf{C}_{\hat{p}}$ dominates over the cosmic variance part $\mathbf{C}_u$. Given enough clusters, the $S/N$ would eventually be limited by $\mathbf{C}_u$, but one can always get around this by including clusters at higher redshift, which provide independent realizations of the uncorrelated polarization signal. 

The cosmic-variance limited $S/N$ (obtained by switching off instrumental noise) is 35.6 for the \emph{Planck} clusters and 27.4 for SPTPol, in the cNFW model. In this case, the variance from the uncorrelated polarization, $p_u$, limits the $S/N$. Since $\emph{Planck}$ has superior sky coverage to SPT, its clusters cover more uncorrelated patches of $p_u$ and hence beat down the cosmic variance to a greater extent. For SPT, this is partly mitigated by the fact that the survey is deep, and high-redshift clusters contribute to higher multipole moments in the signal power spectrum, which can be probed by a smaller field of view.

Because of its finite sky coverage, the signal that SPT can measure will either benefit or suffer from the variation of the polarization field over the sky. To understand this variation, consider a cluster located at $z=0$. The polarization generated is then a projection of the local temperature quadrupole. We may represent the quadrupole using a symmetric trace-free tensor $\mathbf{Q}$, such that $T_2(\vnhat) = \vnhat^{\top} \mathbf{Q}\,\vnhat$. Using the spherical-harmonic expansion of this field, we can express the quadrupole coefficients $a_{2m}$ in terms of the elements of $\mathbf{Q}$. Using the spin-weighted spherical harmonic expansion of Eq.~\eqref{eq:pol_in}, we can express the polarization field in terms of $\mathbf{Q}$\footnote{Alternatively we can use $Q \pm iU = -\tau [\mathbf{Q}]_{ij}\mathrm{e_{\mp}}^i \mathrm{e_{\mp}}^j/10$, where the local complex polarization basis $\ve_{\pm} = \hat{\bm{\theta}} \pm i  \hat{\bm{\phi}}$.}. Following this procedure, it is straightforward to find that the magnitude of the polarization field has two maxima on opposite sides of the sky (the quadrupole has even parity), aligned with the normal to the plane containing the temperature quadrupole maxima and minima (we assume that all the eigenvalues of $\mathbf{Q}$ are different and non-zero, which is true for our Universe). For explicit forms of the quadrupole polarization, see \citep{1999MNRAS.310..765S}. The direction of maximal polarization is an eigenvector of $\mathbf{Q}$, and intersects the unit sphere at $(l,b)=(67.5^{\circ},-67.9^{\circ})$ and $(l,b)=(247.5^{\circ},67.9^{\circ})$ in Galactic coordinates, using the $a_{2m}$ coefficients from the \emph{Planck} SMICA map. For clusters at $z=0$, these would be the `sweet-spots' of the polarization field. These sweet-spots are discernible for the low-redshift source in Fig.~\ref{fig:maps}, although free-streaming has moved the maxima somewhat from their pure-quadrupole locations.

We have seen that scattering at higher redshift transfers power to higher multipoles, as well as generating polarization uncorrelated with the local temperature. By varying the location of the centre of the SPT field across the sky, and keeping the redshift distribution fixed, we find that the maximum of the $S/N$ is at $(262.8^{\circ},-59.4^{\circ})$ in Galactic coordinates. In celestial coordinates the maximum is at $(2^{\mathrm{h}}48^{\mathrm{m}},-48^{\circ})$. Note that there is only one maximum, due to the contribution of odd multipoles that spoils the even parity of the quadrupole field. Centering the SPT field on this direction gives a $S/N$ of 0.32, assuming all other SPT survey parameters are fixed and using the cNFW model. The SPTPol field location therefore provides roughly 90\% of the maximum $S/N$ achievable with the SPT cluster redshift distribution and survey size.

How would the $S/N$ change if we centred on the location of the ACTPol field? Using the ACT sky coverage map shown in Fig.~3 of \citep{2010SPIE.7741E..51N}, the maximum $S/N$ spot should be observable, and lies in a part of the sky already observed with ACT in 2008. We would therefore expect ACT to be able to capture all the $S/N$ available if its redshift distribution is similar to that of SPT. Note that neither the ACT deep fields (D1--D6 used in the recent power spectrum analysis \citep{2014arXiv1405.5524N}, which are all equatorial) or the proposed wide field (which overlaps with BOSS) actually include the polarization `sweet-spot', although it is observable with ACT (and at higher elevation than for SPT).
 
\begin{figure}
\centering
\includegraphics[width=0.6\textwidth]{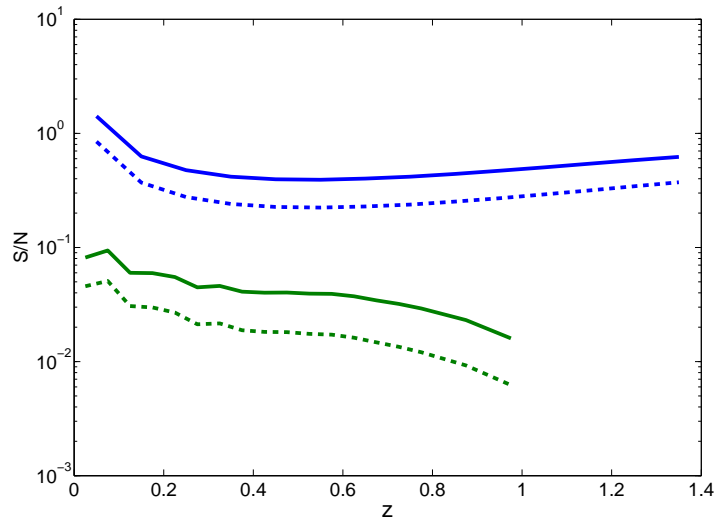}
\caption{(Color Online.) $S/N$ for the \emph{Planck}-like survey with the $\beta$-model (green, lower solid) and cNFW model (green, lower dashed), and SPTPol with the $\beta$-model (blue, upper solid) and cNFW model (blue, upper dashed), assuming that all clusters in each survey are located in a single redshift bin.}
\label{fig:sn_planck_spt_zs}
\end{figure}

We can also investigate the redshift-dependence of our constraints. In Fig.~\ref{fig:sn_planck_spt_zs} we plot the $S/N$ for each experiment and each cluster model, assuming that all clusters in each survey are located in a single redshift bin. The curves in this plots represent a trade-off between the size of the signal and the resolution of the survey. All the clusters in the surveys are approximately the same physical size, so those at lower redshift appear larger and are hence better resolved. For this reason, the $S/N$ increases when all the clusters are at low redshift. At higher redshift, the noise variance increases due to the finite resolution of the beam. This is particularly acute for the \emph{Planck}-like survey, with its $7\,\rm{arcmin}$ resolution, where the $S/N$ decreases significantly with redshift. However SPT can still resolve these high-redshift clusters quite well, and so the $S/N$ actually increases due to the increasing gas profile normalization, which scales as $E(z)^2$. Figure~\ref{fig:sn_planck_spt_zs} makes it clear that SPT could measure the quadrupole polarization with some significance if all the clusters were at low redshift.

Finally, we can consider the improvement made when local $E$-mode CMB polarization is included as a prior in addition to the local temperature. Since the signal is large-scale, we are essentially conditioning our analysis on measurement of the large-scale polarization. We optimistically ignore Galactic foreground emission, assuming this can be cleaned with multi-frequency observations. Using ensemble averages for the local temperature and $E$-mode covariances when calculating the S/N, in the cNFW model, the $S/N$ for the \emph{Planck}-like survey improves by 7\%, and for SPTPol the improvement is 20\%. In the $\beta$-model, the $S/N$ improves by 9\% and 22\% for \emph{Planck} and SPTPol respectively. The gains are marginal due to much of the correlation of the signal with the local $E$-mode coming from high-redshift clusters not present in the surveys (see Fig.~\ref{fig:zetaE}). Note that in obtaining these results, we have neglected the contaminating contribution of the $E$-mode to the observed polarization towards clusters. Hence, we do not include large-scale $E$-mode measurements in the $S/N$ calculation at this stage, since this systematic must be properly accounted for (see Sec.~\ref{sec:E-bias}).

\section{Bias and variance from the kinetic polarization}
\label{sec:K-bias}

In this section we investigate the consequences of ignoring the kinetic part of the polarization signal, sourced by the relative motion of the cluster with respect to the CMB rest-frame. All numerical results in this section assume a cNFW gas profile for the clusters.

Suppose the cluster has transverse velocity components $V_\theta$ and $V_\phi$ with respect to the line of sight\footnote{Note that the components $V_\theta$ and $V_\phi$ are defined with respect to an orthonormal basis.}. The polarization due to the kinetic term in a direction $\vnhat$ observed at frequency $\nu$ is
\begin{equation}
\label{eq:kterm}
(Q \pm iU)(\vnhat) = -\frac{\tau(\vnhat)}{10}T_0\left(\frac{x}{2}\coth {\frac{x}{2}}\right)(V_\theta \mp iV_\phi)^2,
\end{equation}
where $x = h \nu/k_{\mathrm{B}} T_0$ is the dimensionless frequency, $T_0$ is the CMB temperature, and we have set the speed of light $c=1$. Note the distinctive frequency dependence, which can allow this term to be separated from the quadrupole signal.
%\ACtwo{As well as switching to $V_\theta$ and $V_\phi$, I switched the $\pm$ since the l.h.s.\ has spin $\mp 2$ in the line-of-sight description we are using.}

If we neglect the contribution of the kinetic term, our estimate of $\alpha$ from each cluster will be biased. However, since we are averaging over many clusters, these biases cancel to a degree.

Defining $\mathbf{p}_K$ as the vector containing the $\tau$-demodulated kinetic polarization at each cluster location, the contribution to $\hat{\alpha}$ from the kinetic polarization is given by
\begin{equation}
\label{eq:bias}
\hat{\alpha}_{\rm K} = \frac{\operatorname{Re} \,[\mathbf{p}_c^{\dag}(\mathbf{C}_{\hat{p}} + \mathbf{C}_u)^{-1}\mathbf{p}_K]}{\mathbf{p}_c^{\dag}(\mathbf{C}_{\hat{p}} + \mathbf{C}_u)^{-1}\mathbf{p}_c}.
\end{equation}
%
%To proceed, we generate random realizations of the two components of the transverse velocity field, assuming no correlation between them. We assume that these velocity components follow a Gaussian distribution with zero mean and rms velocity $300 \, \km \, \rm{s}^{-1}$. The assumption of a Gaussian distribution should be fairly accurate if the clusters are advected with large-scale bulk flows. % Even in the presence of a non-linear velocity field this assumption is fairly accurate, since by symmetry the lowest-order non-Gaussian term in the velocity distribution is the connected 4-point function, which is suppressed relative to the 2-point function by two factors of $V_x$ (or $V_y$).
%
%We also assume that the velocities are uncorrelated between clusters. For \emph{Planck} this is a reasonable assumption, since for clusters at $z \sim 1$ the correlation length of the velocity field subtends an angle of $\sim 1^{\circ}$, which is below the rms angular separation of clusters in the catalogue. However, the assumption is not so good for SPT, which has a much smaller field of view. Indeed, ACT has recently measured correlated bulk flows in a similar-sized field by using the kinetic SZ effect \citep{2012PhRvL.109d1101H}. A more accurate treatment would need to account for these correlated velocities. 
%
Of course, we do not know the transverse velocities of our cluster sample a priori. If we were to average over realisations of the velocity field, what would be the mean bias in our estimator? Naively, this is given by the ensemble average of $\mathbf{p}_K$, which vanishes upon inspection of Eq.~\eqref{eq:kterm}. However, we must also take into account that the average is conditioned on both the value of our local temperature anisotropies and the cluster number density (see discussion in Sec.~\ref{subsec:MLestimator}).

%quadrupole, and the fact that clusters form in overdense halos. The mean bias is then proportional to the ensemble average of $\mathbf{p}_K$ conditional on the value of the local quadrupole and the halo distribution. The latter is related to the linear matter overdensity by a bias factor, as discussed in Sec.~\ref{sec:like}. The scale-dependence of the terms contributing to the these terms suggests the dominant contribution of the local quadrupole is through the ISW term, where the integration over time allows small spatial scales to contribute at late times. %Note that neglecting either \emph{one} of the overdensity field or the local quadrupole in this procedure still ensures vanishing bias in the mean, since $V_x \pm iV_y$, $\delta_m$ and $p_c$ all have zero mean. For example, neglecting the correlation with the local quadrupole, $\langle \mathbf{p}_K \rangle$ conditional on $\delta_m$ is proportional to $\langle(V_x \pm i V_y)^2 \delta_m \rangle $, which is the product of three zero-mean Gaussian variabes, and hence vanishes.
%

Firstly we will consider the expected bias conditional on the local temperature anisotropies neglecting the fact that clusters form in overdensities for now. We can quantify the magnitude of the bias in a straightforward manner assuming linear perturbation theory\footnote{As we shall see, the bias calculated in linear theory is very small. There may be an additional bias from the squeezed limit of non-linear evolution, whereby large-scale modes that contribute to $\mathbf{p}_c$ modulate the small-scale velocity power (note that the kinetic polarization is quadratic in the cluster velocity).}
To proceed, it will prove useful to expand the spin $\pm 1$ quantity $V_\theta \pm iV_\phi$ in spin spherical harmonics
\begin{equation}
\label{eq:vpm}
(V_\theta \pm iV_\phi)(\vnhat) = \sum_{lm} {}_{\pm}V_{lm}(r) {}_{\pm 1}Y_{lm}(\vnhat).
\end{equation}
We derive the multipole coefficients in Appendix~\ref{app:vlm}; the answer is
\begin{equation}
\label{eq:vlm}
{}_{\pm}V_{lm}(r) = \pm 4\pi i^{l} \sqrt{l(l+1)} \int \frac{\mathrm{d}^3\vk}{(2\pi)^{3/2}} \, v(\vk,r) \frac{j_l(kr)}{kr} Y_{lm}^*(\vkhat),
\end{equation}
where we have written the three-dimensional velocity field as a pure gradient $\mathbf{v} = -\nabla v / k$, since we only consider scalar perturbations. This ensures that the multipole coefficients have electric parity. Note that this is a pure dipole at $r=0$. Linear perturbation theory may now be used to compute the velocity potential from the initial curvature fluctuation. 
%
%\ACtwo{Have made various $\pm$ switches to maintain consistency with Appendix~\ref{app:vlm}.}

We can now calculate the mean values of each multipole coefficient given knowledge of the local CMB temperature anisotropies. This is simply given by the part of ${}_{\pm}V_{lm}$ correlated with the CMB
\begin{equation}
\langle {}_{\pm}V_{lm} | a_{lm} \rangle = \frac{\langle {}_{\pm}V_{lm} a^*_{lm} \rangle}{C_l^{TT}}a_{lm}.
\label{eq:velcorr}
\end{equation}
Note that averaging over realizations of our local CMB results in zero bias, as can be seen from the form of Eq.~\eqref{eq:bias}. Using Eq.~\eqref{eq:velcorr}, we can compute the conditional velocity average, and then use Eq.~\eqref{eq:bias} to compute the bias. There is no bias from the uncorrelated component of the velocity by statistical isotropy, although it does add additional variance (see below). For both \emph{Planck} and SPT, we find that the result is small, of order $10^{-5}$. This is at least partly due to the smallness of the correlation between the CMB and the $V_\theta \pm iV_\phi$, which we find to be roughly 5\% at $l=2$ at $z=0.05$, rising to 30\% at $z=1$. 

There are two reasons for the smallness of the CMB-velocity correlation. Firstly, as mentioned above, the velocity field is a pure dipole at $z=0$, whereas the CMB scales that contribute to the sum in Eq.~\eqref{eq:bias} are purely quadrupolar at $r=0$, giving zero cross-correlation. Thus the bias is suppressed for low-redshift clusters. Secondly, the dominant contribution is from the ISW effect. This contributes most power to $\langle {}_{\pm}V_{lm} a^*_{lm} \rangle$ at wavenumbers between $10^{-4} \, \mathrm{Mpc}^{-1}$ and  $10^{-2} \, \mathrm{Mpc}^{-1}$ for $l=2$--$5$. In contrast, the velocity power spectrum peaks at smaller scales, roughly at $0.05 \, \mathrm{Mpc}^{-1}$, such that the integral over wavenumber is suppressed. The Bessel function in Eq.~\eqref{eq:vlm} ensures a greater overlap in scales at high redshift for low $l$, which explains why the correlation is higher at $z=1$ than at $z=0.05$, but this is somewhat offset by the sharp drop in the velocity power spectrum at these scales.

Finally we note that a marginally better estimate of the velocity field could be made by using all the information contained in the CMB rather than the just the largest angular scales. Small angular scales in $V_\theta \pm iV_\phi$ can survive the correlation with the large-scale polarization term in Eq.~\eqref{eq:bias} since the kinetic polarization is proportional to the velocity \emph{squared}, which results in mode coupling. However, the CMB-velocity correlation has most power on large angular scales for the redshift range under consideration, so we lose little by conditioning on only the large-scale CMB temperature anisotropies.

We now briefly return to the issue that we only sample $p(\vnhat;r)$ at cluster locations, which are a biased tracer of large-scale structure. We shall simply neglect this effect since to generate a significant kinetic bias from further conditioning on the  cluster number density would require a large asymmetry between the components of the velocity field as a result of our given realisation of the density field. To the extent that such an asymmetry is small, we can neglect the effect. In summary, we expect that neglecting the kinetic term should add negligible bias to our estimator. 

%To calculate the mean value of $\langle \mathbf{p}_K|\delta_m \rangle$, we need to know the size of the overdensity at each cluster location, but as discussed in Sec.~\ref{sec:like}, this cannot be extracted accurately from SZ catalogues alone. For these reasons, we will simply neglect this term. This should be fine, since for the value of $\langle (V_x \pm iV_y)^2 | \delta_m \rangle$ to be significantly different from zero would require a significant asymmetry between $V_x$ and $V_y$ to be present in the mean as a result of our given density field realization. To the extent that this is small, we can neglect this contribution. In summary, we expect that neglecting the kinetic term should add negligible bias to our estimator. 

We can also calculate the extra variance in the estimated $\alpha$ due to the kinetic polarization. This amounts to calculating the variance of Eq.~\eqref{eq:bias} assuming its mean vanishes. This procedure is straightforward if we assume that the velocity field is Gaussian distributed, since we can then use Wick's theorem to express the variance in terms of the 2-point function of $(V_\theta \pm iV_\phi)$. The correlation function should be conditioned on our local temperature anisotropies, such that only the \emph{uncorrelated} velocity component adds variance. However, as discussed above, the correlation is weak at the relevant redshifts, so we approximate by using the correlation function of the \emph{total} velocity field. This is given by
\begin{equation}
\langle (\bar{V}_\theta + i\bar{V}_\phi)(\vnhat_1;r)(\bar{V}_\theta \mp i\bar{V}_\phi)(\vnhat_2;r')\rangle = \pm \sum_l \frac{2l + 1}{4\pi} \xi_l^V(r,r') \, d^l_{1\pm1}(\beta),
\end{equation}
where $\langle {}_{+}V_{lm}(r) {}_{+}V^*_{l'm'}(r') \rangle \equiv \xi_l^V(r,r')\delta_{ll'}\delta_{mm'}$, and an overbar denotes the quantity rotated onto the geodesic connecting the two points, as described in Sec.~\ref{sec:physics}. For reference, we find that the r.m.s.\ three-dimensional velocity at $z=0.05$ is roughly $500 \, \mathrm{km}\,\mathrm{s}^{-1}$. We calculate the extra variance in the estimate of $\alpha$ to be roughly $4\times 10^{-5}$ for the \emph{Planck}-like survey and $1\times 10^{-4}$ for SPT, i.e., completely negligible. This may be understood as being due to the disparity in the relevant scales: it is difficult for the velocity field of the cluster sample to mimic a very large scale almost-quadrupolar polarization signal. This is likely the reason why the variance is higher for SPT, since the field of view is smaller such that the coherence scale of the velocity field is a greater fraction of the total observable area. Indeed, when we switch off correlations between cluster velocities, the variance due to the kinetic term falls by 70\% for SPT as compared to 30\% for \emph{Planck}, suggesting large-scale coherent flows are most detrimental to measuring the quadrupole polarization.

Finally we note that any residual bias and variance from the kinetic effect may be mitigated by making use of the distinctive frequency dependence of this signal. It may even be possible to separate out the kinetic term and hence measure the transverse velocity field. Recently, \citep{2013arXiv1306.5248R} have proposed to measure the kinetic polarization signal from local bulk flows, finding that the effect may be measurable for large enough velocities with near-term experiments. We leave these considerations to a future work.

\section{Bias and variance from the background CMB}
\label{sec:E-bias}

A further complication to our analysis comes from the fact that we observe the total polarization towards a cluster, which includes a contribution from the background CMB itself. This term cannot be separated with multi-frequency data since it has the same spectrum as the quadrupole signal.

The background CMB polarization may be expressed in terms of large-scale $E$-modes that are coherent over the cluster, and sub-cluster scale modes. Both of these will bias our measurements if not properly accounted for. 

%However, we have a good understanding of the statistics of CMB polarization. Most of the angular structure in the $E$-mode is at small scales ($l \approx 1000$), corresponding to cluster or sub-cluster scales. We should be able to use measurements of these small scales away from clusters to predict accurately the signal behind each cluster, using standard CMB inpainting techniques.

%This leaves larger super-cluster scale polarization modes, which could in principle contribute a bias to our estimator. We can quantify this bias in a straightforward manner. In harmonic space, the bias from large-scale $E$-modes is

%Firstly consider the large-scale $E$-modes. We should be able to remove these modes using standard CMB inpainting techniques, but the bias can be quantified in a straightforward manner. In harmonic space, the bias is simply
%
%\begin{equation}
%\label{eq:E-bias}
%B(\hat{\alpha}) = \frac{\operatorname{Re}\,(\sum_{lm}p_{c,lm}^*(N_l + C_{u,l})^{-1}E_{lm})}{\sum_{lm}p_{c,lm}^*(N_l + C_{u,l})^{-1}p_{c,lm}},
%\end{equation}
%
%where $N_l$ is the power spectrum of $\mathbf{C}_{\hat{p}}$ etc. Replacing $E_{lm}p_{c,lm}^*$ with its ensemble average, we can evaluate the sum trivially. By inspection, we would expect the bias to be of order $C_2^{TE}/\zeta_2^T \sim 1\%$. Indeed the bias is 0.012 for \emph{Planck}  and 0.016 for SPT, assuming the cNFW profile, i.e. at the percent level.

Firstly consider the large-scale $E$-modes. The bias can be quantified in a straightforward manner, and is given by
\begin{equation}
\label{eq:E-bias}
B(\hat{\alpha}) = \left\langle\frac{\operatorname{Re} \,[\mathbf{p}_c^{\dag}(\mathbf{C}_{\hat{p}} + \mathbf{C}_u)^{-1}\hat{\mathbf{p}}_E]}{\mathbf{p}_c^{\dag}(\mathbf{C}_{\hat{p}} + \mathbf{C}_u)^{-1}\mathbf{p}_c}\right\rangle,
\end{equation}
where the average is conditional on the local temperature anisotropies.
The vector $\hat{\mathbf{p}}_E$ contains the output of the $\tau$-demodulation procedure on the background $E$-mode for each cluster, with two entries per clusters ($Q\pm iU$). For the $j$th cluster, this is given by
\begin{equation}
\hat{p}^{(j)}_E = \frac{\bm{\tau}_{(j)}^{\top}\mathbf{N}^{-1}\mathbf{E}^{(j)}}{\bm{\tau}_{(j)}^{\top}\mathbf{N}^{-1}\bm{\tau}_{(j)}} = E^{(j)}\frac{\sum_i \tau_i}{\sum_i \tau_i^2},
\end{equation}
where $E^{(j)}$ is the background CMB polarization towards the $j$th cluster. We have assumed that these large-scale modes are constant over the cluster such that the vector $\mathbf{E}^{(j)}$, whose elements are the background polarization in each cluster, is given simply by $\mathbf{E}^{(j)} = E^{(j)}\mathbf{t}$, where $\mathbf{t}$ is a vector with each component equal to unity. Note that the expectation value of $\hat{\mathbf{p}}_E$ does not vanish when conditioned on the local CMB temperature.

Evaluating Eq.~\eqref{eq:E-bias} shows that the bias from these modes is significant, equal to 12.8 for the \emph{Planck}-like survey and 20.7 for SPT, assuming the cNFW model. This could have been anticipated, since the background $E$-mode is sourced by scattering of the quadrupole at reionization on these scales. The power in the quadrupole has not changed significantly between $z=10$ and cluster redshifts (see Fig.~\ref{fig:eta_zlag}), so the magnitude of the bias is of order the ratio of the optical depths to reionization and cluster scattering, i.e., of order $10^{-1}/10^{-2} \sim 10$. 
%The background also has a part correlated with the local temperature, and thus possesses angular structure comparable to the signal we are trying to measure. 
As well as adding a bias, large-scale $E$-modes increase the variance on $\hat{\alpha}$, which is given approximately by the variance of the quantity inside the angle brackets in Eq.~\eqref{eq:E-bias}. This degrades the $S/N$ by a factor of roughly 20\% for \emph{Planck}, and almost 90\% for SPT. Note that there is also a small correlation between the signal and the $E$-mode background, which is proportional to $\zeta_l^E$. This should further reduce the extra scatter from the background, since upward fluctuations in background polarization are accompanied by upward fluctuations in the signal since $\zeta_l^E$ is positive.

Fortunately, the large-scale $E$-mode background may be removed from the analysis by the following simple construction. % Including the background CMB, the likelihood of Eq.~\eqref{eq:loglike2} generalises to
% %
% \begin{equation}
% \label{eq:loglikeE}
% -2\log{L(d | p,E)} = \sum_{i,j,k,m}(d_i^{(j)} - \tau_i^{(j)}p^{(j)}-E^{(j)})^{*}(\mathbf{N})^{-1}_{i,(j),k,(m)}(d_k^{(m)} - \tau_k^{(m)} p^{(m)}-E^{(m)}).
% \end{equation}
% %
% Consider marginalising over $E^{(j)}$ on a cluster-by-cluster basis, with a Gaussian prior on $E^{(j)}$ having zero mean and variance $\sigma_{E}$. We will assume that the possible values that can be taken by $E^{(j)}$ are independent between clusters. The result is the likelihood of Eq.~\eqref{eq:loglike2}, but with a noise covariance matrix whose inverse is given by
% %
% \begin{equation}
% \mathbf{N}^{-1} = \left(\sigma_N^2 \mathbf{I} + \sigma_E^2 \mathbf{t}\mathbf{t}^{\top}\right)^{-1} = \sigma_N^{-2}\mathbf{I} - \frac{\sigma_N^{-4}\mathbf{t}\mathbf{t}^{\top}}{\sigma_E^{-2} + \sigma_N^{-2}\mathbf{t}^{\top}\mathbf{t}},
% \end{equation}
% %
% where $\mathbf{I}$ is the identity matrix. Note that this is half the inverse-covariance of $Q$ (or $U$) over the cluster, with $\mathrm{cov}(Q,U) = 0$ for all pixels in a given cluster.
% To remove the constant background, take the limit $\sigma_E \rightarrow \infty$ in this expression.
When demodulating the optical depth across each cluster to construct the cluster polarization signal, we additionally marginalise over a constant background. This is equivalent to replacing the inverse of the pixel noise covariance matrix according to
$\mathbf{N}^{-1} \rightarrow \sigma_N^{-2}(\mathbf{I} - \hat{\mathbf{t}}\hat{\mathbf{t}}^{\top})$ where $\hat{\mathbf{t}} = \mathbf{t}/|\mathbf{t}|$. The estimator of the projected quadrupole at the cluster is then
\begin{equation}
\label{eq:phat_marg}
\hat{p} = \frac{\bm{\tau}^{\top}(\mathbf{I} - \hat{\mathbf{t}}\hat{\mathbf{t}}^{\top})\mathbf{d}}{\bm{\tau}^{\top}(\mathbf{I} - \hat{\mathbf{t}}\hat{\mathbf{t}}^{\top})\bm{\tau}} = \frac{1}{N_{\mathrm{pix}}}\left[\frac{\sum_i (\tau_i -\bar{\tau})d_i}{\overline{\tau^2} - \bar{\tau}^2}\right],
\end{equation}
where $\bar{\tau}$ and $\overline{\tau^2}$ are respectively the mean and mean-square optical depth of the cluster, and we have dropped the cluster label for clarity. The estimator Eq.~\eqref{eq:phat_marg} has variance
\begin{equation}
\label{eq:phat_var_marg}
\mathrm{var}(\hat{p}) = \frac{\sigma_N^2}{N_{\mathrm{pix}}\left(\overline{\tau^2} - \bar{\tau}^2\right)},
\end{equation}
where $N_{\mathrm{pix}}$ is the number of pixels in the cluster field. Note that the variance now receives a non-zero contribution from pixels outside the cluster profile, since these still carry information on the constant background mode. In the limit of infinitely many such pixels, we would know the background mode precisely and hence recover the no-background expression Eq.~\eqref{eq:phat_var}. 

The projection operator $\mathbf{I} - \hat{\mathbf{t}}\hat{\mathbf{t}}^{\top}$ in Eq.~\eqref{eq:phat_marg} removes any constant mode in the data, and hence removes the bias from background large-scale polarization, by construction. The price we pay for this is increased variance in the estimator. We find that the $S/N$ on $\hat{\alpha}$ degrades by roughly 30\% for both the \emph{Planck}-like survey and SPT, assuming the cNFW model, and using only pixels where $\tau \ne 0$. The degradation is mild, demonstrating the usefulness of this approach. For the $\beta$-model, the $S/N$ falls by a factor of about 40\% for \emph{Planck} and 50\% for SPT. From Fig.~\ref{fig:tau} we see that the optical depth profile is much flatter in the $\beta$-model, giving more variance since the signal has less contrast against the background. Note that we can mitigate the drop in $S/N$ by including pixels around the cluster where $\tau = 0$, as discussed above.

In contrast, small-scale $E$-modes contribute negligible bias, since they do not survive the correlation with the large-scale CMB temperature in Eq.~\eqref{eq:E-bias}. The extra variance from these modes is also very small, giving negligible degradation in the $S/N$ for both \emph{Planck} and SPT. The small residual bias may be removed by marginalising over these modes, which adds extra variance to the projected quadrupole estimator Eq.~\eqref{eq:phat}\footnote{Equation~\eqref{eq:phat_varE} ignores the background subtraction, and assumes that instrument noise dominates over the small-scale $E$ modes. Note, further, that for circularly-symmetric optical depth profiles, the generalised noise in the two complex components of $\hat{p}$ remains uncorrelated.}:
%
%We can also evaluate the extra cosmic variance incurred by marginalising over the large-scale background $E$-modes. This amounts to evaluating the denominator of Eq.~\eqref{eq:E-bias} with an extra noise term given by $C_l^{EE}$. The reduction is $S/N$ is at the sub-percent level for both surveys, due to the dominant source of variance coming from the experimental noise and finite resolution.
%
%Next we must consider small-scale modes. Most of the angular structure in $E$-modes is at these small scales ($l \approx 1000$), corresponding to cluster or sub-cluster scales. Assuming that these modes are uncorrelated between clusters, we can calculate the extra variance that these modes contribute by including an extra term in Eq.~\eqref{eq:loglike2}
%
%\begin{equation}
%\label{eq:loglikeE}
%-2\log{L(d | p,E)} = \sum_{i,j,k,m}(d_i^{(j)} - \tau_i^{(j)}p^{(j)}-E_i^{(j)})^{*}(\mathbf{N})^{-1}_{i,(j),k,(m)}(d_k^{(m)} - \tau_k^{(m)} p^{(m)}-E_k^{(m)}),
%\end{equation}
%
%where $E_i^{(j)}$ is the (beam-smoothed) background CMB polarization in $i$-th pixel of the $j$-th cluster. Since our estimator for the projected quadrupole Eq.~\eqref{eq:phat} is linear in the data, the CMB background gives no bias, but does induce extra variance, given by 
%
\begin{equation}
\label{eq:phat_varE}
\mathrm{var}(\hat{p}) =  \frac{\sigma_N^2}{\sum_i \tau_i^2} + \frac{\sum_{ik}\tau_i \tau_k \langle E_i E_k^* \rangle}{(\sum_i \tau_i^2)^2}.
\end{equation}
%
% \fixme{This seems to assume you are doing no background removal. Moreover, it is only correct for low signal-to-noise on the primordial polarization. Marginalising over the small-scale modes (with no background projection) just replaces $N_{ij}$ everywhere with $\sigma_N^2 \delta_{ij} + \langle E_i E^*_j \rangle$, so
% %
% \begin{equation}
% \frac{1}{\rm{var}(\hat{p})}= \sum_{ij} \tau_i \tau_j [\sigma_N^2 \delta_{ij} + \langle E_i E^*_j \rangle ]^{-1} \, .
% \end{equation}
% %
% In the limit of low signal-to-noise on $E$ (i.e., $\sigma_N^2 \gg \rm{var}(E)$), this reduces to Eq.~\eqref{eq:phat_varE}.} 

The extra term is readily calculable in Fourier space. Using Parseval's theorem, the numerator is given approximately by
\begin{equation}
\label{eq:Enum}
\sum_{ik}\tau_i \tau_k \langle E_i E_k^* \rangle \approx \frac{1}{2\pi \Omega_{\mathrm{pix}}^2}\int^{\infty}_{l_{\mathrm{min}}} \mathrm{d}l \,  [\tilde{\tau}(l)]^2 l C_l^{EE} e^{-l^2 \sigma_b^2},
\end{equation}
%
% Note that
%
%\begin{equation*}
%\langle (Q+iU)(\vx)(Q-iU)(\vx') \rangle = \int \frac{d^2 \vl}{(2\pi)^2} C_l^{EE}
%e^{i\vl\cdot(\vx -\vx')} \, ,
%\end{equation*}
%
%and the Fourier convention claimed for $\tilde{\tau}(\vl)$ below.}
where $\Omega_{\mathrm{pix}}$ is the area of each pixel, and $\sigma_b^2$ is the variance of the beam function, which we take to be Gaussian. Our Fourier transform convention is
\begin{equation}
\label{eq:fourier}
\tilde{\tau}(\mathbf{l}) = \int \mathrm{d}^2\mathbf{x}\, \tau(\mathbf{x}) e^{i \mathbf{l} \cdot \mathbf{x}}.
\end{equation}
We impose a cut-off $l_{\mathrm{min}}$ on the Fourier sum given by the inverse angular radius of the cluster, such that only sub-cluster modes are included. For both the \emph{Planck}-like survey and SPT, we find that the extra variance is negligible in comparison to the instrumental noise term, by several orders of magnitude and for both gas models. The reason for this is that for low-redshift clusters the optical depth profile has angular structure on larger scales than the background CMB, which suppresses the contribution of the second term in Eq.~\eqref{eq:phat_varE}. For the high-redshift clusters, the relevant scales are comparable, but the oscillatory structure of the CMB power spectrum suppresses the contribution of the integral in Eq.~\eqref{eq:Enum} relative to the noise term, where the equivalent integrand is much smoother in $l$.

\section{High-redshift estimator}
\label{sec:hi-z}

The estimator, Eq.~\eqref{eq:alpha_hat}, is based on an approximation to the conditional likelihood [Eq.~\eqref{eq:loglikefull}] appropriate at low redshift. 
However, for a high-redshift survey, one may become concerned that too much information is being thrown away since only the correlation of the cluster polarization with the local temperature anisotropies is considered. There is additional information in the uncorrelated component, $p_u$, that may enhance detection.

To capture this extra information we consider the full $\alpha$ dependence in the likelihood~\eqref{eq:loglikefull}. Demodulating the optical depth as in Sec.~\ref{subsec:opticaldepth}, we have
\begin{equation}
\label{eq:loglike_hiz}
-2\ln{L(\hat{p} | \alpha, p_c)} = (\hat{\mathbf{p}} - \alpha \mathbf{p}_c)^{\dag}(\mathbf{C}_{\hat{p}} + \alpha^2 \mathbf{C}_u)^{-1}(\hat{\mathbf{p}} - \alpha \mathbf{p}_c) + \ln{\det(\mathbf{C}_{\hat{p}} + \alpha^2 \mathbf{C}_u)}.
\end{equation}
Minimising this with respect to $\alpha$ requires us to solve an equation of the form $\mathcal{M}(\alpha) - \langle \mathcal{M}(\alpha) \rangle = 0$. Following \citep{2009PhRvD..80f3004H}, (see also \citep{1998PhRvD..57.2117B}), we can solve this iteratively using Newton's method:
\begin{equation}
\label{eq:newton}
\alpha_{i+1} = \alpha_i - \frac{\mathcal{M} - \langle \mathcal{M} \rangle}{d\left(\mathcal{M} - \langle \mathcal{M} \rangle\right)/d\alpha} \bigg|_{\alpha=\alpha_i}.
\end{equation}
A sensible choice for the starting point is $\alpha_0 = 1$. Approximating the denominator in Eq.~\eqref{eq:newton} by its mean value, we find the next-order solution
\begin{multline}
\label{eq:hiz_est}
\hat{\alpha} = 1+\frac{1}{\mathcal{F}_{\alpha\alpha}}\left\{
(\hat{\mathbf{p}}-\mathbf{p}_c)^{\dag}(\mathbf{C}_{\hat{p}}+\mathbf{C}_u)^{-1}\mathbf{C}_u(\mathbf{C}_{\hat{p}}+\mathbf{C}_u)^{-1}(\hat{\mathbf{p}}-\mathbf{p}_c) - \mathrm{Tr} \left [\mathbf{C}_u(\mathbf{C}_{\hat{p}}+\mathbf{C}_u)^{-1} \right] \right.
\\
\left. + \operatorname{Re} \left[ (\hat{\mathbf{p}}-\mathbf{p}_c)^{\dag}(\mathbf{C}_{\hat{p}}+\mathbf{C}_u)^{-1}\mathbf{p}_c \right] \right\} ,
\end{multline}
where the Fisher information on $\alpha$ is
\begin{equation}
\label{eq:fisher}
\mathcal{F_{\alpha \alpha}} = \mathbf{p}_c^{\dag}(\mathbf{C}_{\hat{p}}+\mathbf{C}_u)^{-1}\mathbf{p}_c + 2\mathrm{Tr} \left [ \mathbf{C}_u(\mathbf{C}_{\hat{p}}+\mathbf{C}_u)^{-1}\mathbf{C}_u(\mathbf{C}_{\hat{p}}+\mathbf{C}_u)^{-1} \right ].
\end{equation}
Whilst this estimator only maximises the likelihood asymptotically, it is straightforward to check that it is unbiased and minimum-variance, in the sense that the Cramer--Rao bound is saturated, $\mathrm{Var}(\hat{\alpha}) = \mathcal{F}_{\alpha \alpha}^{-1}$. It also has the advantage of being a quadratic estimator, making it readily calculable. The Fisher information is a sum of two positive terms, one coming from the information in the correlated component of the cluster polarization and the other from the power in the uncorrelated component\footnote{%
The estimator in Eq.~\eqref{eq:hiz_est} for $\hat{\alpha}-1$ can be written in the form
\begin{equation}
\hat{\alpha}-1 = \frac{1}{\mathcal{F}_{\alpha\alpha}} \left[\mathcal{F}_{\alpha\alpha}^{\rm corr}
(\hat{\alpha}-1) + \frac{1}{2} \mathcal{F}_{\alpha^2 \alpha^2}^{\rm uncorr} \left(\widehat{\alpha^2}-1\right) \right] ,
\end{equation}
where $\hat{\alpha}^{\rm corr}$ is the estimator in Eq.~\eqref{eq:alpha_hat}, with Fisher information $\mathcal{F}_{\alpha\alpha}^{\rm corr}$ given by the inverse of Eq.~\eqref{eq:sn}, and
\begin{equation}
\widehat{\alpha^2} = \frac{1}{2 \mathcal{F}^{\rm uncorr}_{\alpha^2\alpha^2}} \left[
(\mathbf{d}-\mathbf{p}_c)^{\dag}(\mathbf{C}_{\hat{p}}+\mathbf{C}_u)^{-1}\mathbf{C}_u(\mathbf{C}_{\hat{p}}+\mathbf{C}_u)^{-1}(\mathbf{d}-\mathbf{p}_c) - \mathrm{Tr}\left[
\mathbf{C}_u(\mathbf{C}_{\hat{p}}+\mathbf{C}_u)^{-1}\mathbf{C}_{\hat{p}}(\mathbf{C}_{\hat{p}}+\mathbf{C}_u)^{-1} \right] \right] ,
\end{equation}
with Fisher information $\mathcal{F}^{\rm uncorr}_{\alpha^2\alpha^2}$ given by one quarter of the second term on the right of Eq.~\eqref{eq:fisher}. The estimator $\widehat{\alpha^2}$ is the usual quadratic maximum-likelihood estimator~\cite{1997PhRvD..55.5895T} for the amplitude of the \emph{power} $\alpha^2$ of the uncorrelated polarization. The construction $(\widehat{\alpha^2}-1)/2$ gives a first-order estimate of the deviation $\alpha-1$, and we see that the estimator for $\hat{\alpha}-1$ combines this information with that in the correlated polarization with inverse-variance weighting.}.

We can compute Eq.~\eqref{eq:fisher} to forecast the $S/N$ on the high-redshift estimator. The only modification to our previous $S/N$ expression is the additive trace term. We find that the $S/N$ for \emph{Planck} changes negligibly, and increases by only 0.1\% for SPT. This modest increase reflects the relative magnitudes of $\mathbf{C}_u$ and $\mathbf{C}_{\hat{p}}$.

Further insight into the form of the estimator in Eq.~\eqref{eq:hiz_est} can be gained by considering some exceptional cases. In the limit that two clusters are in close proximity, the estimator is a quadratic function of the inverse-noise-weighted combination of $(\hat{\mathbf{p}}_1$ and $\hat{\mathbf{p}}_2)$. This makes sense intuitively, since the clusters are located in the same quadrupole, so averaging the data and then squaring beats down the noise most effectively. In the opposite limit of no correlation, data from the two clusters decouple and the estimator is quadratic in $\hat{\mathbf{p}}_1$ and $\hat{\mathbf{p}}_2$ separately.

We would expect the quadratic estimator to be optimal for a high-redshift survey. However, we have shown that the simpler linear estimator gives comparable S/N, implying negligible loss of information for low-redshift SZ surveys.

\section{Conclusions and future prospects}
\label{sec:conc}

The polarization from scattering the remote temperature quadrupole in galaxy clusters is too weak to detect in individual clusters with current and upcoming surveys. A statistical approach combining many clusters is therefore necessary. We have argued that for low redshift clusters, the best prospect for detection is from that part of the cluster polarization that is correlated with the local, large-angle temperature and polarization anisotropies. We constructed a linear estimator to measure optimally the correlated signal, and forecast the $S/N$ for a full-sky \emph{Planck}-like survey, and a ground-based survey similar to SPTPol. The detection significance is quite sensitive to the optical depth profile, which we assume is known for each cluster. For the cNFW profile, we find that a \emph{Planck}-like survey can only achieve a $S/N$ of 0.03, assuming the clusters observed are the same as those currently in the SZ catalogue. For SPTPol the $S/N$ is 0.28. (A similar number is forecast for ACTPol.)

We also considered biases induced by the kinetic part of the polarization signal, and the background $E$-mode which is spectrally indistinguishable from the quadrupole signal. The bias from the kinetic term was found to be negligible for the surveys we considered, due to the disparity of the relevant scales. We also note that the distinctive frequency dependence of the kinetic signal may allow for the transverse velocity to be directly measured by such a survey, a consideration left for a future work. Similarly, we found that both the bias and variance from small-scale background $E$-modes are small, due to the contrasting scale-dependence of these modes and the large-scale correlated polarization. However large-scale $E$-modes introduce significant bias and variance into the analysis because of the weakness of the cluster polarization and the similar angular dependence of the cluster and background signals. Fortunately, under the assumption that these modes are coherent across each cluster, they can be projected out of the estimator with only a mild degradation in the $S/N$. 

Our results suggest the correlated polarization signal is undetectable with the \emph{Planck}, SPTPol and ACTPol cluster surveys. The survey and experimental parameters for the small-scale CMB experiments are comparable, but the ACT is better located to observe the maximum of the correlated polarization (determined approximately by the orientation of the local quadrupole) at high elevation. Despite these conclusions, it should be emphasised that these $S/N$ forecasts may be significantly improved by increasing the integration time on each cluster. For example, neglecting the cosmic variance of the uncorrelated polarization, a $2\sigma$ detection of the correlated polarization may be made with SPTPol if the noise variance were reduced by a factor of roughly 50. This could be achieved, for example, by increasing the integration time on each cluster by a factor of 50 with dedicated follow-up with a telescope having the noise specifications of SPTPol.

We have neglected the impact of polarized emission from sources intrinsic to the cluster itself. Typical sources include weakly polarized synchrotron radiation from diffuse radio halos in the central regions of clusters, and highly polarized emission from radio relics in the cluster outskirts \citep{2012A&ARv..20...54F}. Since the angle of polarization should be uncorrelated between clusters, averaging over clusters as our optimal estimator prescribes will result in a cancellation and the estimator will remain unbiased. The extra variance contributed will depend on both the variance of the intrinsic polarization in each cluster, and its spatial coherence scale compared to the beam size. Recent MHD simulations of radio halos in \citep{2013A&A...554A.102G} indicate an r.m.s. polarization of roughly $3 \times 10^{-7} \, \mathrm{Jy} \, \mathrm{arcsec}^{-2}$ at 1.4 GHz. Assuming a power-law index of $-3$ for the energy distribution of relativistic electrons, this translates into a brightness temperature of roughly $0.05 \, \mu \K$ at 150 GHz and $z=0$. The contribution from radio halos is thus subdominant to the detector noise in a pixel for both \emph{Planck} and SPTPol. Even if the intrinsic polarization were a significant contamination, as may be the case for clusters containing radio relics or well-resolved clusters observed with future SZ surveys, its distinctive synchrotron spectrum should enable its removal with multi-frequency observations.

We note further that our results have been obtained using only a single frequency channel for each experiment. A realistic survey would have to use several channels to remove foregrounds, but we do not expect our $S/N$ forecasts to change significantly due to the poorer resolution and sensitivity to polarization offered by the other channels in each experiment.

Although the situation is not optimistic for upcoming surveys, next-generation polarization experiments with broad frequency coverage are being planned, such as PRISM \citep{2013arXiv1306.2259P}. This experiment should detect roughly $10^6$ clusters, with many thousands at $z>2$. With the excellent sensitivity, full sky coverage and large cluster sample, we would expect a definite detection to be made. Using forecast PRISM detector sensitivities from \citep{2013arXiv1306.2259P}, with $10^6$ clusters and the \emph{Planck} SZ survey redshift distribution, a $S/N$ of 6.1 is achievable with just the 160 GHz channel, assuming the cNFW gas model. With its excellent frequency coverage, we would expect PRISM to outperform this estimate. The formalism we have laid out in this work will be important for stacking these clusters in the optimal way, especially since the survey depth is greater and hence the relative importance of the uncorrelated signal is larger.

Going beyond a first detection of cluster polarization, it is interesting to ask whether this new observable will allow for improved constraints on cosmological models (see also~\cite{2006PhRvD..73l3517B,2004PhRvD..70f3504P}). Forming the joint likelihood for the local temperature anisotropies and the cluster polarization, by multiplying the likelihood in Eq.~\eqref{eq:loglike_hiz} by the likelihood for the local temperature, we can calculate the Fisher matrix for cosmological parameters $\mu$ and $\nu$. For the simple case of a full-sky, isotropic survey (see Appendix~\ref{app:harm}) at a single redshift, the Fisher matrix reduces to
\begin{equation}
F_{\mu\nu} = F_{\mu\nu}^T + \sum_l (2l+1) \frac{C_l^{TT}}{C_l^{\hat{p}}+C_l^u}
\frac{\partial}{\partial \mu} \left(\frac{\zeta_l^T}{C_l^{TT}}\right)
\frac{\partial}{\partial \nu} \left(\frac{\zeta_l^T}{C_l^{TT}}\right)
+ \sum_l \frac{2l+1}{2} \frac{1}{(C_l^{\hat{p}}+C_l^u)^2} \frac{\partial C_l^u}{\partial \mu}
\frac{\partial C_l^u}{\partial \nu} ,
\end{equation}
where $C_l^u = \xi_l - (\zeta_l^T)^2/C_l^{TT}$ is the power spectrum of the uncorrelated polarization and $C_l^{\hat{p}}$ is the noise power spectrum after demodulating the cluster optical depth. The Fisher matrix $F_{\mu\nu}^{T}$ is the information from the local temperature alone. The remaining two positive-definite terms are the information from the correlated polarization and the uncorrelated polarization. The latter is limited by the poor signal-to-noise, but in principle does carry valuable additional information on $\rm{Gpc}$-scale modes~\cite{2006PhRvD..73l3517B}. For the correlated cluster polarization, no extra information is added on parameters that affect $\zeta_l^T$ and $C_l^{TT}$ in the same way, such as the amplitude of the primordial power spectrum $A_s$. However, for parameters that affect the two spectra differently, such as dark energy or the shape of the primordial power spectrum, the correlated polarization \emph{does} add information. (A familiar example is constraints on the optical depth to reionization from the $TE$ correlation.) However, even for low-redshift surveys for which the correlation is high [$(\zeta_l^T)^2 \approx \xi_l C_l^{TT}$], the information is strongly suppressed by the low signal-to-noise on the auto-spectrum $\xi_l/C_l^{\hat{p}}$.
Although the correlated part of the cluster polarization will bring little extra information, the point we have stressed in this paper is that it will provide a very useful cross-check on the $\Lambda \mathrm{CDM}$ model. Since this makes a specific prediction for the amplitude of cluster polarization, any significant deviation from this could be indicative of new physics. Dedicated observations with current SZ survey instruments or a future space-based mission will allow for such a measurement.

Finally, we note that our signal calculation has neglected relativistic corrections and finite temperature effects, which could be important for some of the hotter clusters in the catalogues \citep{2000MNRAS.312..159C}. In particular, both the kinetic signal and the quadrupole signal receive corrections proportional to the ICM temperature and the CMB rest-frame cluster velocity, albeit with distinctive frequency dependencies. In principle the thermal terms can be corrected for because the gas temperature can be extracted from the X-ray measurements required to derive the optical depth profile. The relativistic term proportional to the velocity is a sub-dominant contribution, but is potentially relevant to a next-generation survey like PRISM.

\section{Acknowledgments}
We thank Richard Battye, Camille Bonvin, Steven Gratton, Yvette Perrott, Debora Sijacki and Stephen Walker for useful correspondence and conversations. AH was supported by an Isaac Newton Studentship from the University of Cambridge, and the Isle of Man Government. Some of the results in this paper have been derived using the \textsc{HEALPix} package \citep{2005ApJ...622..759G}.

\bibliography{references}

\appendix

\section{Harmonic-space $S/N$}
\label{app:harm}

In this appendix we present a simple harmonic-space derivation of our linear estimator for the amplitude of the correlated polarization, Eq~\eqref{eq:alpha_hat}. Firstly, assume that all clusters are at a single redshift. The observed polarization in a pixel $i$ is
\begin{equation}
d_i = \tau_i \, p_i + n_i,
\end{equation}
where $p_i$ is the projected quadrupole at the location of the cluster containing pixel $i$ with optical depth $\tau_i$, and $n_i$ is the thermal noise in that pixel. If pixel $i$ is in direction $\vnhat_i$, expanding $p_i$ in spin $-2$ spherical harmonics gives
\begin{equation}
d_i = \tau_i \sum_{lm} p_{lm} \, {}_{-2}Y_{lm}(\vnhat_i) + n_i = \tau_i \, p_A \, {}_{-2}Y_A(\vnhat_i) + n_i,
\end{equation}
where we have grouped the multipole indices into a single bulk index $A$, over which summation is implicit. It is convenient to introduce the matrix $\mathbf{M}$, which is a vertical concatenation of matrices $(\mathbf{M}_{\mp})_{iA} = \tau_i \, {}_{\mp 2}Y_{A}(\vnhat_i)$. The maximum-likelihood estimator for $p_{A}$ is then
\begin{equation}
\hat{\mathbf{p}} = [\mathbf{M}^{\dag}(\mathbf{N}\oplus\mathbf{N})^{-1}\mathbf{M}]^{-1}\mathbf{M}^{\dag}(\mathbf{N}\oplus\mathbf{N})^{-1}\mathbf{d},
\end{equation}
which has covariance
\begin{equation}
\label{eq:harmcov}
\langle \Delta \hat{\mathbf{p}} \Delta \hat{\mathbf{p}}^{\dag} \rangle = [\mathbf{M}^{\dag}(\mathbf{N}\oplus\mathbf{N})^{-1}\mathbf{M}]^{-1},
\end{equation}
where the direct sum $\mathbf{N}\oplus\mathbf{N}$ is the full noise covariance matrix and $\Delta \hat{\mathbf{p}} = \hat{\mathbf{p}} - \langle \hat{\mathbf{p}} \rangle$. Note that $\mathbf{d}$ includes both $Q+iU$ and $Q-iU$ for each cluster. The matrix $\mathbf{M}^{\dag}(\mathbf{N}\oplus\mathbf{N})^{-1}\mathbf{M}$ decomposes as $(\mathbf{M}_{-}^{\dag} \mathbf{N}^{-1} \mathbf{M}_{-})\oplus(\mathbf{M}_+^{\dag} \mathbf{N}^{-1} \mathbf{M}_+)$, with terms
\begin{equation}
(\mathbf{M}_{\mp}^{\dag}\mathbf{N}^{-1}\mathbf{M}_{\mp})_{lm,l'm'} =\left[\sum_j {}_{\mp2}Y^*_{lm}(\vnhat_j){}_{\mp2}Y_{l'm'}(\vnhat_j)\sum_{i \in j} \frac{\tau_i^2}{\sigma_N^2}\right],
\end{equation}
where we have split the sum into a piece over clusters $j$ and a piece over pixels $i$ within each cluster. This split is justified because the quadrupole coherence length is assumed to be much larger than the size of the cluster, so the spin harmonics can be taken as constant over the cluster. Note that this matrix can have full rank even for a handful of clusters if the maximum $l$ is chosen to be sufficiently small. 

We will assume that there are $N$ clusters at the given redshift, each having the same optical depth profile. For a survey with full sky coverage, we can approximate the sum over clusters by
\begin{equation}
\sum_j^N {}_{-2}Y_{lm}(\vnhat_j){}_{-2}Y^*_{l'm'}(\vnhat_j) \approx \frac{1}{\Omega} \int \mathrm{d}^2\vnhat\, {}_{-2}Y_{lm}(\vnhat){}_{-2}Y^*_{l'm'}(\vnhat) = \frac{1}{\Omega} \delta_{ll'} \delta_{mm'},
\end{equation}
where $\Omega = 4\pi/N$. The covariance matrix Eq.~\eqref{eq:harmcov} is hence diagonal. It is now straightforward to derive the harmonic space analogue of the $\alpha$ estimator in Eq.~\eqref{eq:alpha_hat}. The variance of $\hat{\alpha}$ is
\begin{equation}
\label{eq:vara}
\mathrm{var}(\hat{\alpha}) = \left( \sum_{lm}  \frac{p_{c,lm} \, p^*_{c,lm}}{C^{\hat{p}} + C^u_l} \right)^{-1},
\end{equation}
where $p_{c,lm}$ are the multipole moments of the correlated polarization, $C^{\hat{p}} $ are the diagonal entries of the covariance matrix in Eq.~\eqref{eq:harmcov}, and $C^u_l$ is the angular power spectrum of the uncorrelated polarization. These formulae are very accurate for the \emph{Planck}-like survey, which has full sky coverage, but underestimate the $S/N$ by about 30\% for SPT, which has finite sky coverage. 

Finally, we note that the extension to multiple redshift bins is straightforward. The term in brackets in Eq.~\eqref{eq:vara} is replaced by a quadratic form with the $p_{c,lm}$ factors promoted to vectors and the $(C^{\hat{p}} + C_l^u)^{-1}$ term replaced by the inverse covariance matrix between redshift bins.

\section{Harmonic expansion of the transverse velocity field}
\label{app:vlm}

In this appendix we compute the multipole coefficients of the transverse velocity field, useful for analysing the kinetic polarization signal. We will denote the transverse velocity field by $\mathbf{V}_{\perp}$, which is a vector on the 2-sphere, and hence may be decomposed into gradient and curl parts as
\begin{equation}
\label{eq:vperp}
\mathbf{V}_{\perp}^a = \nabla^a \Omega + \epsilon^a_{\; \, b} \nabla^b \Psi,
\end{equation}
where $\nabla$ is the covariant derivative on the sphere, and $\epsilon^a_{\; \, b}$ is the alternating tensor. We introduce the spin raising and lowering operators $\eth$ and $\bar{\eth}$ respectively, which act on a spin $s$ quantity ${}_{s}\eta$ as
\begin{align}
\label{eq:eth}
\eth {}_{s} \eta = -\sin^{s}\theta (\partial_{\theta} + i \, \mathrm{cosec}\, \theta \, \partial_{\phi})(\sin^{-s}\theta \, {}_{s}\eta),\\
\bar{\eth} {}_{s} \eta = -\sin^{-s}\theta (\partial_{\theta} - i \, \mathrm{cosec}\, \theta \, \partial_{\phi})(\sin^{s}\theta \, {}_{s}\eta).
\end{align}
We define the spin $\pm 1$ harmonics in term of the standard spherical harmonics by acting once with the spin operators
\begin{align}
\label{eq:spindef}
{}_{1}Y_{lm} = \eth \left(\sqrt{\frac{(l-1)!}{(l+1)!}} Y_{lm} \right),\\
{}_{-1}Y_{lm} = -\bar{\eth} \left(\sqrt{\frac{(l-1)!}{(l+1)!}} Y_{lm} \right).
\end{align}
The quantity of interest for the kinetic polarization is the transverse velocity projected onto the null basis $\hat{\bm{\theta}} \pm i \hat{\bm{\phi}}$. Using Eqs.~\eqref{eq:vperp} and \eqref{eq:eth} we have
\begin{align}
\label{eq:ethphi}
V_{\theta} + iV_{\phi} = -\eth (\Omega - i\Psi),\\
V_{\theta} - iV_{\phi} = -\bar{\eth}(\Omega + i\Psi).
\end{align}
Now expand the potentials $\Omega$ and $\Psi$ in spherical harmonics
\begin{align}
\Omega(\vnhat) = \sum_{lm} \sqrt{\frac{(l-1)!}{(l+1)!}}\epsilon_{lm} Y_{lm}(\vnhat),\\
\Psi(\vnhat) = \sum_{lm} \sqrt{\frac{(l-1)!}{(l+1)!}}\beta_{lm} Y_{lm}(\vnhat),
\end{align}
where $l \geq 1$. Using Eqs.~\eqref{eq:ethphi} and \eqref{eq:spindef} then gives
\begin{equation}
(V_{\theta} \pm i V_{\phi})(\vnhat) = \sum_{lm}(\mp \epsilon_{lm} + i\beta_{lm}){}_{\pm 1}Y_{lm}(\vnhat).
\end{equation}
Under a parity transformation, the velocity field transforms as $\mathbf{v}(\vnhat) \rightarrow -\mathbf{v}(-\vnhat)$, such that the transverse components transform as $(V_{\theta} \pm iV_{\phi})(\vnhat) \rightarrow -(V_{\theta} \mp iV_{\phi})(-\vnhat)$. This implies that the multipole coefficients transform as $\epsilon_{lm} \rightarrow (-1)^l \epsilon_{lm}$ and $\beta_{lm} \rightarrow (-1)^{l+1}\beta_{lm}$, such that the gradient and curl terms have electric and magnetic parity respectively.

Scalar perturbations imply that only the gradient term is present. In Fourier space, we have $\mathbf{v}(\vk) = -i\vkhat v(\vk)$, where we have suppressed the time dependence. In real space the velocity is
\begin{equation}
\mathbf{v}(\bm{r}) = \nabla_{\bm{r}} \int \frac{\mathrm{d}^3\vk}{(2\pi)^{3/2}} \, \frac{-v(\vk)}{k} \, e^{i \vk \cdot \bm{r}},
\end{equation}
where $\nabla_{\bm{r}}$ is the three-dimensional derivative. We now project this onto the 2-space perpendicular to the line-of-sight direction $\vnhat$, and set $\bm{r} = r\vnhat$. The projected three-dimensional derivative is related to the covariant derivative on the sphere by $\nabla^{\perp}_{\bm{r}} = (1/r)\nabla$. The potential $\Omega$ is then given by
\begin{equation}
\Omega(\vnhat, r) = - \int \frac{\mathrm{d}^3\vk}{(2\pi)^{3/2}} \, \frac{v(\vk,r)}{kr}  \, e^{i kr \vkhat \cdot \vnhat}.
\end{equation}
Expanding the exponential in spherical harmonics then gives the multipole coefficients $\epsilon_{lm}$ as
\begin{equation}
\epsilon_{lm}(r) = -4\pi i^l \sqrt{l(l+1)} \int \frac{\mathrm{d}^3\vk}{(2\pi)^{3/2}} \, v(\vk,r) \frac{j_l(kr)}{kr} Y_{lm}^*(\vkhat).
\end{equation}
%
% Finally we note that the same expression may be derived using the methods in Appendix 2 of \citep{2006PhRvD..73l3517B}, upon using the identity
% %
% \begin{equation}
% \vkhat \cdot (\hat{\bm{\theta}} \pm i \hat{\bm{\phi}}) = \mp \sqrt{2} \, \frac{4\pi}{3} \sum_{m=-1}^1 {}_{\pm 1}Y_{1m}(\vnhat) \, Y_{1m}^*(\vkhat),
% \end{equation}
% %
% which is easily checked.

\end{document}